\documentclass[manuscript]{aastex}

\shorttitle{HST Spectral Mapping of Brown Dwarfs}
\shortauthors{Apai et al.}

\begin{document}

\title{HST Spectral Mapping of L/T Transition Brown Dwarfs Reveals Cloud Thickness Variations}

\author{D\'aniel Apai\altaffilmark{1}}
\affil{Department of Astronomy, The University of Arizona, 933 N. Cherry Avenue, Tucson, AZ 85721}
\email{apai@as.arizona.edu}

\author{Jacqueline Radigan\altaffilmark{2}}
\affil{Department of Astronomy, University of Toronto, 50 St. George Street, Toronto, M5S 3H4, Canada}

\author{Esther Buenzli}
\affil{Department of Astronomy and Steward Observatory, 933 N. Cherry Avenue, The University of Arizona,
Tucson, AZ 85721}

\author{Adam Burrows}
\affil{Department of Astrophysical Sciences, 105 Peyton Hall, Princeton University, Princeton, NJ 08544}

\author{Iain Neill Reid}
\affil{Space Telescope Science Institute, 3700 San Martin Drive, Baltimore, MD 21212}

\and

\author{Ray Jayawardhana}
\affil{Department of Astronomy, University of Toronto, 50 St. George Street, Toronto, M5S 3H4, Canada}

\altaffiltext{1}{Department of Planetary Sciences, 1629 E. University Blvd, Tucson, AZ 85721}
\altaffiltext{2}{Space Telescope Science Institute, 3700 San Martin Drive, Baltimore, MD 21212}


\newcommand{\changes}{}

\begin{abstract}
Most directly imaged giant exoplanets are fainter than brown dwarfs with similar spectra. To explain their relative underluminosity unusually cloudy atmospheres have been proposed. However, with multiple parameters varying between any two objects, it remained difficult to observationally test this idea. We present a new method, sensitive time-resolved Hubble Space Telescope near-infrared spectroscopy, to study two rotating L/T transition brown dwarfs (2M2139 and SIMP0136). The observations provide spatially and spectrally resolved mapping of the cloud decks of the brown dwarfs. The data allow the study of cloud structure variations while other parameters are unchanged. We find that both brown dwarfs display variations of identical nature: J- and H-band brightness variations with minimal color and spectral changes. Our light curve models show that even the simplest surface brightness distributions require at least three elliptical spots. We show that for each source the spectral changes can be reproduced with a linear combination of only two different spectra, i.e. the entire surface is covered by two distinct types of regions. Modeling the color changes and spectral variations together reveal patchy cloud covers consisting of a spatially heterogenous mix of low-brightness, low-temperature thick clouds and brighter, thin and warm clouds.
We show that the same thick cloud patches seen in our varying brown dwarf targets, if extended to the entire photosphere, predict near-infrared colors/magnitudes matching the range occupied by the directly imaged exoplanets that are cooler and less luminous than brown dwarfs with similar spectral types. This supports the models in which thick clouds are responsible for the near infrared properties of these "underluminous" exoplanets.

\end{abstract}

\keywords{stars: individual (MASS J21392676+0220226, 2MASS J0136565+093347) --- stars: low-mass, brown dwarfs --- stars: spots --- stars: atmospheres --- planets and satellites: individual (Jupiter) --- stars: planetary systems }


\section{Introduction}

With masses between cool stars and giant exoplanets and effective temperatures comparable to those of directly imaged exoplanets \citep[e.g.][]{Chauvin2005,Marois2008,Lafreniere2008,Marois2010,Lagrange2010,Skemer2011} L and T-type brown dwarfs provide the critical reference points for understanding the atmospheres of exoplanets \citep[e.g.][]{Burrows2001,Kirkpatrick2005,Marley2007}. Because the observations of brown dwarfs are not limited by the extreme star-to-planet contrasts exoplanet observations pose, much more detailed studies can be carried out. In particular, brown dwarfs provide an opportunity to solve the puzzling observation that most directly imaged giant planets appear to be redder and up to 4--10 times fainter than typical brown dwarfs with the same spectral type \citep[e.g.][]{Barman2011_HR8799,Skemer2012}, often referred to as the {\changes {\em under-luminosity problem}. Particularly interesting well-studied examples are see in Ross 458C \citep[][]{Burgasser2010,Burningham2011,Morley2012} and 2M1207b \cite{Chauvin2005,Mohanty2007,Patience2010,Barman2011_2M1207,Skemer2011}. Although for the case of \object{2M1207b} an obscuring edge-on disk with grey extinction has been proposed as a solution, in the light of additional observations and analysis this solution appears very unlikely \citep[][]{Skemer2011}. More likely is that the fainter and redder near-infrared emission is due to a property intrinsic to the atmospheres of these exoplanets. This possibility is further supported by the fact that similar underluminosity has also been reported for a handful of field brown dwarfs (e.g. \object{HD 203030B}: \citealt[][]{Metchev2006}, \object{HN Peg B}: \citealt[][]{Luhman2007}, \object{2MASSJ18212815+1414010}, \object{2MASSJ21481628+4003593}: \citealt[][]{Looper2008}) and young brown dwarfs in clusters \citep[][]{Lucas2001,Allers2006}. }
The different models proposed to explain the lower near-infrared luminosity of exoplanets and {\changes brown dwarfs} invoke differences in elemental abundances, surface gravity differences, evolutionary state, chemical equilibrium/non-equilibrium, or cloud structure, or some combination of these. However, because several of these parameters may change between any two brown {\changes dwarfs or exoplanets} it remained difficult to isolate
the effect of these variables. Possible differences in the structure of condensate clouds have, in particular, received much attention in models (e.g. \citealt[][]{AckermanMarley2001,Burgasser2002,Skemer2012,Barman2011_HR8799,Barman2011_2M1207,Burrows2006,Madhu2011,Marley2010} and progress has been made in spectroscopic modeling to separate or constrain the impact of cloud structure from other parameters \citep[e.g.][]{Cruz2007,Folkes2007,Looper2008,Burgasser2008,Radigan2008,Cushing2010}. Yet, this problem remains a challenging aspect
of ultracool atmospheres and one which will benefit from observational data probing cloud properties more directly.

We present here high-cadence, high-precision time-resolved HST spectroscopy of two rotating early T-type brown dwarfs that reveal highly heterogeneous cloud covers across their {\changes photospheres}. These observations allow us to separate the effects of different cloud structures from variations in surface gravity, elemental abundances, age and evolutionary state. We show that the observed variations are well reproduced by models with large cloud scale height variations (thin and thick clouds) across the surfaces. When thick clouds turn to the visible hemispheres both targets fade in the near-infrared and display changes consistent with the colors and brightness of ÒunderluminousÓ directly imaged exoplanets. The similarity of the changes observed provides strong support to models that invoke atmospheres with high dust scale heights to explain the photometry of directly imaged exoplanets. 

\section{Observations and Data Reduction}

\subsection{Observations and Targets}

We used the Hubble Space Telescope (HST) to obtain near-infrared grism spectroscopy of two L/T transition brown dwarfs as part of a larger campaign (Programs 12314, 12551 PI: Apai). 
The data were acquired with the sensitive Wide Field Camera 3 instrument  \citep[][]{MacKenty2010} by obtaining 256$\times$256 pixel images of the targets' spectra dispersed by the G141 grism in six consecutive HST orbits. 
{\changes Table~\ref{ObsLog} provides a log of the observations. In short, we obtained 660 spectra for 2M2139, each with 22.34s integration time; and 495 spectra for SIMP0136, each with 22.34s integration time. In the analysis that follows we averaged sets of 10 spectra for 2M2139 and sets of 5 spectra for SIMP0136, giving us an effective temporal resolution of 223s for 2M2139 and 112s for SIMP0136.
Our targets are relatively bright (J=13.5 mag for SIMP0136 and J=15.3 mag for 2M2139), resulting in very high signal-to-noise spectra (see Sect.~\ref{Uncertainties} for a detailed assessment).}

At the beginning of each orbit a direct image was obtained to accurately determine the position of the source on the detector, required for precise wavelength calibration. Cross-correlation of the images and centroid positions on-source revealed positional differences less than 0.1 pixel (0.01") between images taken at different orbits. No dithering was applied to stabilize source positions, and improve the accuracy of relative measurements. 

\begin{table}
\begin{center}
\caption{Log of observations and noise measurements.\label{ObsLog}}
\begin{tabular}{lccccccc}
\tableline\tableline
Target   & Date & Time & \# of Int. & \#Orbits & Total & Bin & Noise\\ 
Name   &   & Per Int. &  / Orbit &  & \#Spectra & Size & per Int. \\
\tableline
2M2139     &  2010/10/21               &  22.34 s     & 11              & 6            &    660  & 10 & 0.27\% \\
SIMP0136 &  2011/10/10                 &  22.34 s     & 16 or 17     & 6            &   495 & 5  & 0.11\% \\
\tableline
\end{tabular}
\end{center}
\end{table}

\begin{figure}
\begin{center}
\includegraphics[angle=0,scale=0.70]{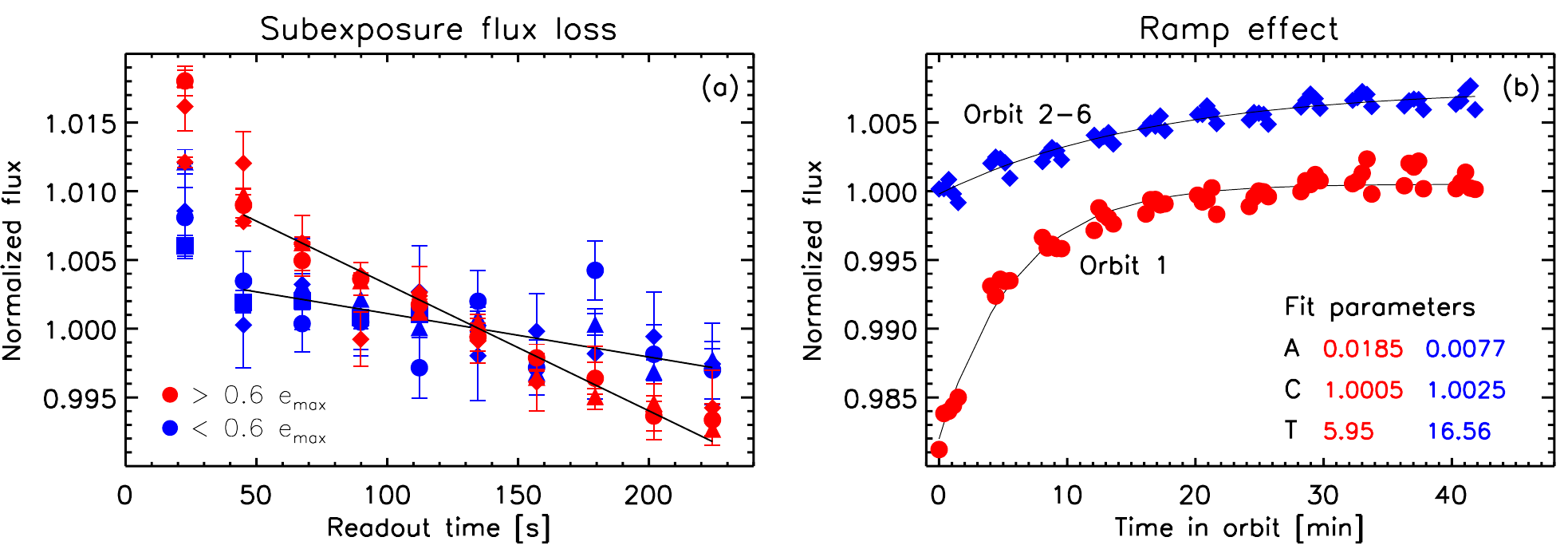}
\caption{Systematic effects observed and corrected in the WFC3 data: flux loss (a) and ramp (b). Both effects are well fitted and removed by simple analytical functions. In (a) sources are shown with counts $<$0.6 (blue) or $>$0.6 (red) of the maximum count in the spectrum. In (b) blue symbols are ramp in orbits 2-6 and red symbols are the ramp in orbit 1. Here the source is a non-variable field star. \label{FigCorrections}}
\end{center}
\end{figure}

The observations presented here focus on two L/T transition brown dwarfs. Target 2M2139 (or 2MASS J21392676+0220226) has been classified {\changes as a T0 dwarf based on red optical spectrum \citep[][]{Reid2008} and as a peculiar T2.5$\pm$1 dwarf based on a 0.8--2.5~$\mu$m spectrum \citep[][]{Burgasser2006}. More recently \citep[][]{Burgasser2010} found that the spectrum of 2M2139 is better fit by a composite spectrum of an earlier (L8.5) and a later type (T3.5) dwarf than any single template brown dwarf. It was recently found to show impressive periodic photometric variability with a peak-to-peak amplitude of $\simeq$27\% \citep[][]{Radigan2012}. Ground-based photometry of variation 2M2139 argues for a period of $7.721\pm0.005$~hr, but also leaves
open the possibility of a two times longer period \citep[][]{Radigan2012}. Recent observations by \citet[][]{Khandrika2013} confirm the variability and argue against a double-peak period.
Based on JHK light curves and a fit to the spectrum
by \citep[][]{Burgasser2006} these authors argue that cloud thickness variations are likely responsible for the photometric
variations seen in 2M2139.}
Target SIMP0136 (2MASS J0136565+093347), another T2 brown dwarf, has also been reported variable \citep[][]{Artigau2009}.  These targets are the first L/T transition sources observed in our two ongoing HST surveys; further results, including coordinated HST/Spitzer observations and sources with later spectral types are discussed in \citet[][]{Buenzli2012} and other upcoming papers.

\subsection{Data Reduction}

Our reduction pipeline combined the standard aXe pipeline with {\changes a } custom-made IDL script, which included corrections for different low-level detector systematics, critical for highly precise relative spectroscopy. We used two-dimensional spectral images from the standard WFC3 pipeline, which were bias and dark current-subtracted, and corrected for non-linearity and gain. 
Bad pixels were marked by a corresponding flag in the data quality plane. In order to correct for detector systematics we started with the {\tt .ima} files, which contain all non-destructive sub-reads of each exposure, rather than the combined {\tt .flt} images. Because our observations were not taken with a dithering pattern, we did not use the standard pipeline's {\tt MultiDrizzle} routine.

We first extracted all sub-reads, discarded the first two zero-reads (0 s and 0.27~s), and compared the count rates of the individual pixels over the sub-reads of an exposure to identify and remove outliers by replacing them with the median value. We also corrected flagged bad pixels by interpolating over adjacent good pixels in the same row. 

We identified a systematic nearly linear flux loss from the first to last subread of an exposure, with a steeper slope for the brighter pixels of a spectrum. We empirically determined a slope of $-$0.2\% per 22.34~s (one subread) for pixels of brightness $>$60\% of the maximum in the spectrum, and only $-$0.001\% for pixels below that level (Fig. \ref{FigCorrections}). {\changes The uncertainties of the slopes are negligible, but a significant `zig-zag' pattern is present with a scatter of $\sim0.1\%$. The standard deviation of the normalized fluxes measured at the same exposure but in different orbits is $\leq$0.2\%. }
This relation held for all of our objects, regardless of the absolute brightness of the spectra. Only the first subread had systematically higher flux than expected from the linear relation and had to be corrected individually for each source.

The spectral extraction was executed with aXe \citep[][]{axe2011}. First, the sub-array images were embedded in larger, full frame-sized images to allow aXe to use standard instrument calibration frames, which are full-frame sized. The data quality flag was used to exclude the extra surrounding pixels of the extended frame from the actual data analysis.
As a first step, the aXe pipeline subtracted a scaled master sky frame, with the scale factor determined individually for each imageÕs background level (i.e. excluding the observed spectra). Then, the location of the target spectrum and the wavelength calibration were determined from the direct (non-grism) images. For the source 2M2139, only one direct image was obtained at the beginning of the first orbit and this image was used for each subsequent spectra. 
For SIMP0136, a separate direct image was taken at the beginning of each orbit and used for all spectra in the given orbit. We fixed the spectral extraction width within each orbit, but allowed it to vary between subsequent orbits. The extraction width was determined by summing up all spectra in a single orbit and then collapsing the sum into a one-dimensional vertical profile. The extraction width was then chosen as three times the full-width-half-maximum (FWHM) of a gaussian fit to that profile. This resulted in an extraction width with mean and standard deviation for the six orbits of 6.50$\pm$0.02 for SIMP0136 and 6.56$\pm$0.01~px for 2M2139, respectively. 

In the final step the spectra from each image were flat fielded, extracted, and collapsed using the standard pixel extraction tables of aXe and flux-calibrated with the latest instrument sensitivity curves. This led to one-dimensional spectra with a spectral resolution of R=$\lambda/\Delta\lambda\simeq130$ and highly reliable data over the wavelength range from 1.1 to 1.7 $\mu$m. We calculated the uncertainties as composed of photon shot noise, read-noise, and sky noise. 
A second systematic detector effect became evident at this point. During each orbit, there was a small increase in flux in the form of an exponential ramp (Fig. \ref{FigCorrections}). Because the intrinsic variability of the sources prevented a direct quantification of this effect, we used the 
partial spectrum of a bright star visible in one of our target fields.
The ramp was found to be independent of object brightness. We integrated the stellar spectrum over the full wavelength range and removed exposures where saturation had occurred. We fitted the exponential ramp $C\times(1-Ae^{-t/T})$ to the light curve, where $t$ is the time since the beginning of an orbit. The ramp was very similar for all the orbits between the second and sixth, which we therefore averaged before fitting, but it was a different, stronger effect in the first orbit of a visit, which was fitted and corrected individually. In the final step we corrected the spectra of all our objects by dividing the time-dependent data by the value of the analytical ramp function sampled at the times of the sub-reads. Fig.~\ref{FigCorrections} shows the ramp and its {\changes best-fit parameters. The latter are the following: For Orbit 1 $A=0.0185\pm0.0013$, $T=5.95\pm0.88$, and $C=1.0005\pm0.0005$, while for orbits 2-6 :  $A=0.0077\pm0.0009$, $T=16.65\pm6.02$, $C=1.0025\pm0.0009$. Note, that these uncertainties include the propagated uncertainties from the slope correction described above. The combined uncertainties for the ramp correction lead to a 1$\sigma$ uncertainty of $\simeq$0.15\%.  }

\subsection{Uncertainties}
\label{Uncertainties}

{\changes

In the following we briefly discuss the uncertainties of our measurements. We distinguish three different uncertainties that affect our data: random (white) noise emerging from photon noise and readout noise, systematic wavelength-dependent trends, and systematic time-dependent trends. As explained below, due the fact that our targets are very bright by HST's standards the photon noise is very small (typically well below 0.1\%) and the systematic wavelength-dependent trends are negligible, the residual time-dependent trends dominate the noise in our data.
We characterize the amplitude of each of these three components based on our data.

\subsubsection{White noise}
Random (white) noise is present in our data due to the combination of photon noise, residual dark noise, and read-out noise. While all three components
are present at very low levels, often negligible for practical purposes, we use our data to measure their combined amplitude. To do this we 
extract pixel-to-pixel variations helped by the fact that our temporal resolution ($<$1 minute) significantly exceeds the timescale on which the
astrophysical changes occur. 
We started from the binned spectral cubes containing 66 spectra (each with 10 binned readouts) for 2M2139 and 99 spectra (each with 5 binned samples) for SIMP0136. To measure the white noise we removed the correlated components by first subtracting a 2-pixel-smoothed version of the data, leaving only variations smaller than 2 resolution elements. In the low-resolution HST data most of
the changes even on this small scale are correlated physical changes (i.e. high-frequency residuals of actual spectral features). These features 
are the same in every spectra and can be removed via the subtraction of the median spectrum of the spectral cube. 
This procedure has removed correlated changes in wavelength and in time, leaving us with white noise-dominated data. To measure the white noise amplitude we calculated the standard deviation of the data in each spectrum and their median.
We find that for 2M2139 the 1$\sigma$ noise per resolution element is 0.27\%, while for the brighter SIMP0136 the 1$\sigma$ noise per resolution elements is 0.11\%.

We point out that most of the conclusions in this paper are drawn from brightness variations measured in broad photometric bands, further decreasing 
the importance of white noise. For example, the J-band light curves in our spectra typically contain 55 data points, leading to a white noise contribution
of less than 0.04\%. }


\subsubsection{Time-dependent trends}

{\changes
Because we study temporal changes in our targets it is important to assess the level of time-dependent trends (red noise) in our
data.  Any red noise present in the data would come for a time-dependent sensitivity variations, potentially introduced either by
drifts in the positions of the sources or by sensitivity changes in the instrument. Our measurements of the positions of the sources
show that the targets have remained in precisely the same positions, thus contribution from the former noise factor can not be significant.
The second noise source, changes in the instrumental sensitivity, however, must be characterized through our data.
Due to observing efficiency considerations our observations were taken in  a subarray mode, which has a relatively small 
field of view and thus and do not contain any other sources of comparable brightness to the targets. Therefore, we do not have 
other non-varying sources in the same datasets that could be used to measure time-dependent trends in our data. 
Instead, we use a third brown dwarf target observed identically to our targets to assess time-dependent trends.
This third target, \object{2MASSJ09153413+0422045} (in the following 2M0915), did not show 
variability above the 0.6\% level and thus provides us with a good reference for measuring 
the temporal stability of the observations. 2M0915 is a resolved binary brown dwarf \citep[L7+L7,][]{Reid2006}, allowing us
to measure flux levels for the two sources simultaneously. We point out that although the wings of the two 
spectra show some overlap, this overlap should not affect the photometric stability of the measurements 
and the following assessment of the systematic uncertainties.

We reduced the 2M0915 data set in the same way as data from the other two sources, with the exception
that a larger aperture was adopted to include both of the slightly overlapping sources. The data 
reduction was repeated two times, with the aperture once centered on component A and once centered
on component B.

For the two components we measured a mean standard deviation of points {\em within} the same orbit of 0.16\% and 0.13\%, 
fully consistent with the combined uncertainties of the correction of the systematic effects and the white noise components. The
fact that this standard deviation is not larger demonstrates that there is no measurable systematic trend left uncorrected on timescales
of an orbit or less.

To assess the photometric stability of HST over timescales of multiple orbits we determined the scatter of the mean values in each orbit.
Thus, we calculated the standard deviation of the mean values of the fluxes measured in each of the six orbits. We found that these values
were 0.25\% and 0.13\% for the two components; these values are our estimates for the uncertainties of the photometric stability of HST
over six orbits. 

Thus, based on the measurements of the noise properties we conclude the following: 1) the random
(white) noise level is 0.3\% per resolution element and integration; 2) the random noise of band-integrated
light curves is less than 0.04\% (practically negligible); 3) the photometric stability between orbits is
about 0.25\% or less; 4) photometric trends within an orbit ($<$50 minutes) amount to 0.16\% or less.
Therefore, although the differences discussed in our paper are small (typically few percent levels) they
are all high significance level detections.}

\section{Results}

\begin{figure}
\begin{center}
\includegraphics[angle=0,scale=0.55]{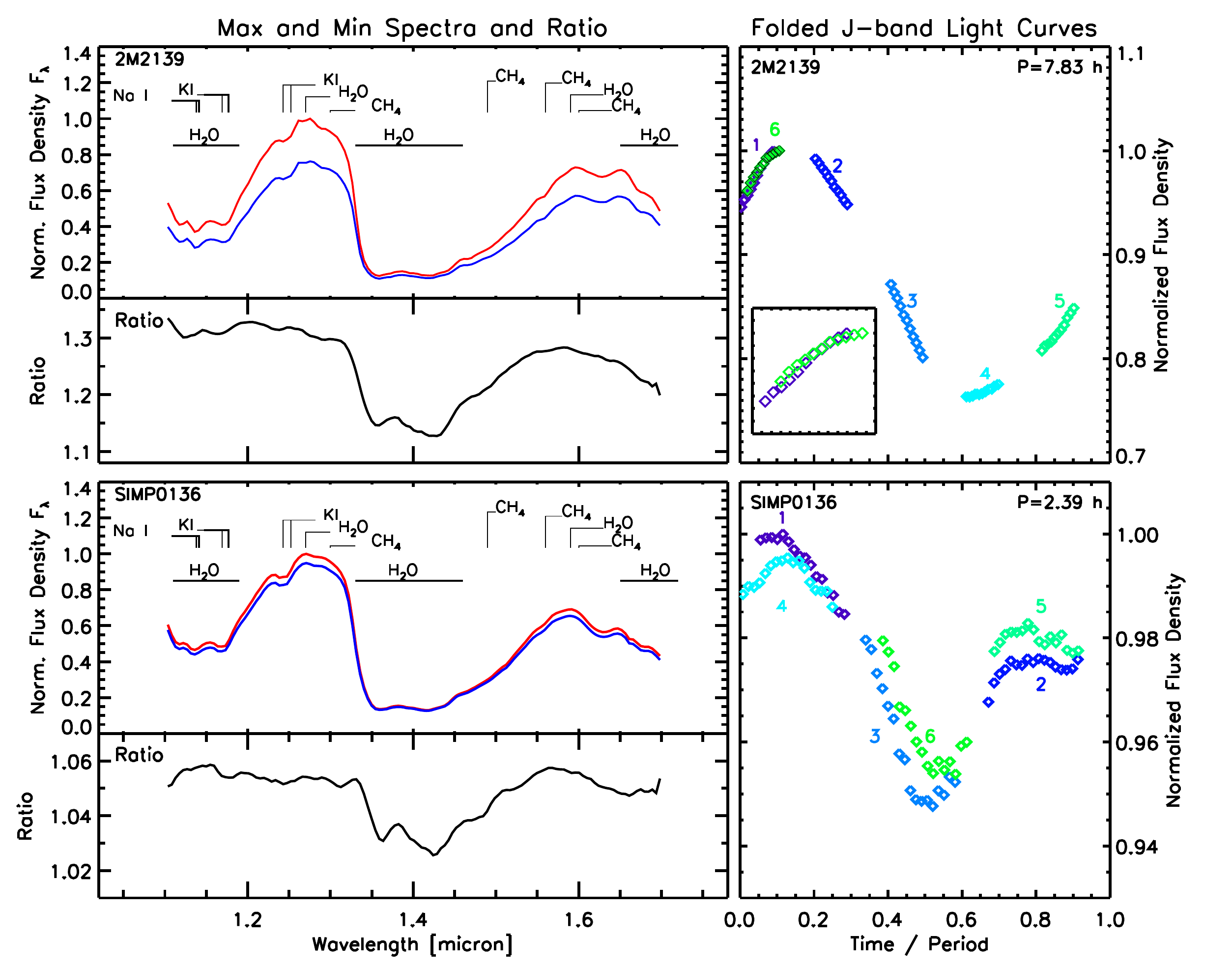}
\caption{Spectra at the faintest and brightest stages of the two brown dwarfs show prominent water, potassium, and methane absorption features with similar depths. The ratio of the minimum  over maximum spectra (minor panels on left) show variations with weak wavelength-dependence in the continuum. and in the potassium, sodium and methane features, but demonstrate lower-amplitude variations in the 1.4 $\mu$m water band. The period-folded J-band light curves (right) reveal variations in the {\changes surface brightness} distributions of these two targets. Red and black colors show data from the first and sixth orbit for 2M2139, which perfectly overlap if a 0.5\% flux scaling is allowed, {\changes consistent with the photometric stability on a 2$\sigma$ level}. In contrast, SIMP0136 displays light curve evolution over 5 hours {\changes present both in the absolute levels and the light curve shape at levels well above our uncertainties}. \label{FigSpectraLC}}
\end{center}
\end{figure}

Our observations provided a series of very high signal-to-noise (SNR$>$300) spectra (see Fig.~\ref{FigSpectraLC}) of the targets covering the 1.1 to 1.7 $\mu$m wavelength range. These spectra probe the J and H broad-band photometric bands, prominent molecular absorption bands (water, methane), as well as atomic resonance lines (K~{\sc I}, Na~{\sc I}). 

The spectra of both targets are dominated by deep and broad water vapor absorption at 1.1 $\mu$m, 1.4 $\mu$m, and 1.7 $\mu$m. Characteristic of L/T transition dwarfs, they both display narrower neutral atomic lines and weaker CH$_4$ absorption. Both sources showed dramatic brightness changes during the observations. Synthetic photometry derived in the core of the standard J- and H-bands (centered at 1.2 and 1.6 $\mu$m) display variations with peak-to-peak amplitudes of 27\% (2M2139) and 4.5\% (SIMP0136). The variations are periodic and we estimate the periods to be 7.83$\pm$0.1~hr and 2.39~$\pm$0.05~hr for 2M2139 and SIMP0136, respectively. 
The period for 2M2139 was derived by least square minimization of the light curve segments overlapping in phase (see inset in Fig.~\ref{FigSpectraLC}) and leads to a 0.35\% standard deviation in the overlapping region, but a slightly imperfect match in light curve shape.   
In contrast, a somewhat shorter period (7.76 h) provides a near-perfect match for the light curve shape, but leads to a slightly higher standard deviation (0.55\%). Given the information at hand we conservatively assume that the period is 7.83$\pm$0.1~h. 
Determining the period for SIMP0136 poses a different challenge: Although the overlap in phase is much larger than it is for 2M2139, the light curve shows a clear evolution during the extent of our observations. We find that assuming a period of 2.39~h aligns the troughs in the light curve, but leads to a shift between consecutive peaks; in turn, a period of 2.42~h aligns the peaks well but leads to
a mismatch in the troughs. This behavior is fully consistent with the light curve evolution observed in this source; in the following we conservatively assume that the period of SIMP~0136 is 2.39~$\pm$0.07h.

These rotational periods were determined by minimizing the differences in phase-folded light curves. The periods are consistent with those reported for these sources from ground-based photometry \citep[][]{Radigan2012,Artigau2009}. For 2M2139 our first and sixth HST orbits cover the same phase for an assumed
period of 7.83~hr. During this overlap the spectra and the flux levels provide very close match {\changes in the shape of the light curve. The flux levels differ by only 0.5\%, which is twice the 1$\sigma$ uncertainty we estimated for the photometric stability of our measurements over multi-orbit timescales (see Section~\ref{Uncertainties}). The fact that the overlapping light curve segments are so similar both in flux and in shape argues} against a twice longer period, a possibility ground-based data left open \citep[][]{Radigan2012}. The 2.39~hr and the 7.83~hr rotations are shorter than Jupiter's rotation period ($\sim$9.92h). 

The left panel of Fig.~\ref{FigSpectraLC} displays the maximum and minimum spectra observed for both targets as well as their ratio (for clarity we do not plot the entire spectral series). The data are of superb quality (S/N$>$300) and allow detailed analysis of the changes. Both targets show a strikingly similar pattern: only weakly wavelength-dependent {\changes broadband} variations. {\changes For the precise shape of the variations we refer the reader to the {\em Ratio} panels in Fig.~\ref{FigSpectraLC} and here only highlight the peak of the ratios in the J and H bands. For 2M2139  the flux density change $\Delta F$ in the observed spectra peaks in the J-band at $33\%$ at 1.20~$\mu$m and in the H-band at a level of 28\% at $1.58~\mu$m. For SIMP0136  $\Delta F$ peaks in the J-band at ${1.15\mu m}$ at 5.9\% and $\Delta F_{1.56\mu m}$ at a level of $5.8\%$. While the changes in the J and H bands are smooth and similar, in both sources the water absorption bands between $\sim1.32-1.50~\mu$m vary at much lower levels: for example $\Delta F_{1.4\mu m}=14\%$ for 2M2139 and $\Delta F_{1.4\mu m}=3.2\%$ for SIMP0136. Note, that given
our uncertainties of 0.25\% for multi-orbit photometric stability (see Sect.~\ref{Uncertainties}) these differences are all highly significant.} 
Surprisingly, with the exception of the water all other gas-phase absorption bands (CH$_{4}$, Na {\sc I}, K~{\sc I}) change together with the continuum. The light curve of 2M2139 shows nearly sinusoidal variations, but SIMP0136 displays sharper, more structured variations. 

\subsection{Spectral Variations and PCA Analysis}
\label{PCA}

Both light curves contain distinct and prominent higher-frequency {\changes components}. We interpret these variations as {\changes spots} with different spectra rotating in and out of the visible hemispheres of the targets. We use the detailed data sets to identify the spectra and spatial distribution of these spots and contrast this information with predictions of state-of-the-art atmosphere models.

\begin{figure}
\includegraphics[angle=0,width=6.5 in]{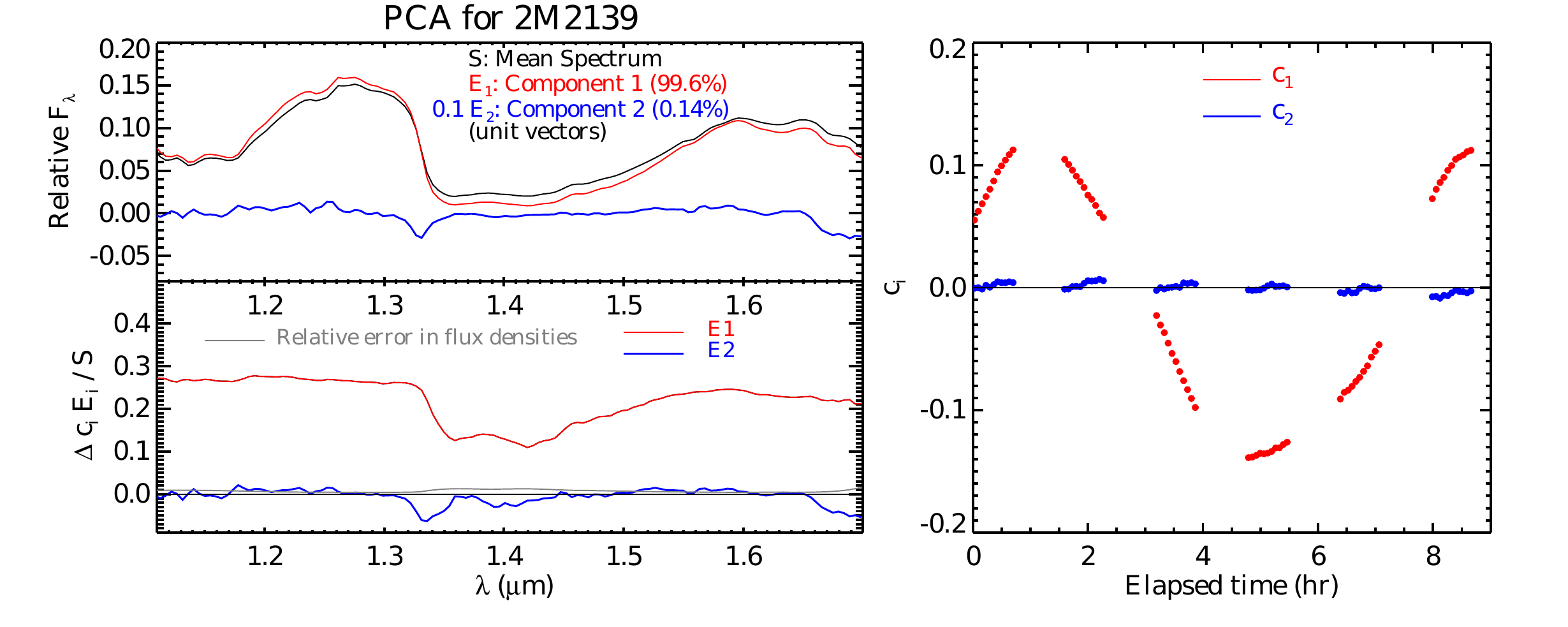}
\includegraphics[angle=0,width=6.5in]{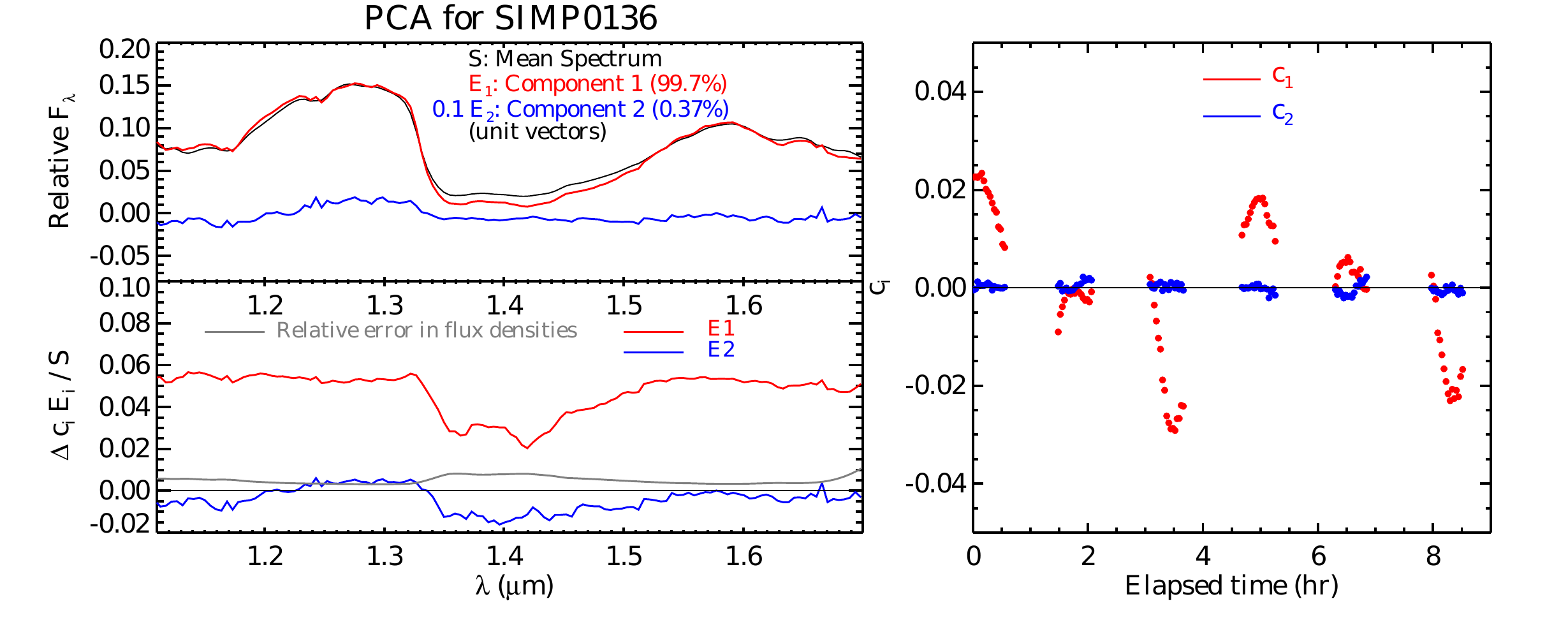}
\caption{Principal Component Analysis of the time series spectra for 2M2139 ({\em top}) and SIMP0136. {\em Top left:} The mean spectrum (black line) and first two principal components of the variability (red and blue lines respectively).  All components have been normalized as unit vectors.  The contributions of each component to the total variability are indicated.  {\em Bottom left:} The principal components plotted relative to the mean spectrum, multiplied by the maximum difference in their time-projections, $\Delta c_i$.  In other words, this panel shows the variability amplitude as a function of wavelength for isolated components.  The relative error in flux densities is shown as a grey line for comparison.  {\em Right}: Projections of the principal components onto the data spectra as a function of time.  The first component is dominant, producing a light curve that mirrors the broadband variations, while the second component appears to cycle with the HST orbits and may reflect low-level uncorrected systematic errors.  \label{FigJackiePCAcomps}}
\end{figure}

{\changes    
We identify the smallest set of independent spectra, over the mean spectrum, that account for the majority of the observed variance by applying a principal components analysis (PCA).  We computed the covariance matrix of the spectral time series over wavelengths of 1.1$-$1.7~$\mu$m, cutting off lower signal-to-noise regions outside this range.  Eigenvectors ({\bf E$_i$}) and eigenvalues ($\Lambda_i$) of the covariance matrix were determined using the {\tt LA\_PACK} routine {\tt LA\_EIGENQL} in $IDL$.  Components were sorted by eigenvalue, and the fractional contribution of each component to the overall variability determined as $\Lambda_i/ (\sum_i \Lambda_i$), where the denominator is a sum over all eigenvalues.   

Every observed spectrum at a given time, ${\bf S}(t)$ can then be approximated by a linear combination of the principal components,  

\begin{equation}
{\bf S}(t) \approx \langle {\bf S}\rangle + c_0(t){\bf E}_0  + c_1(t){\bf E}_1 + ...
\end{equation}

where the series is truncated to include only components that contribute significantly above the noise level.  The coefficients $c_i(t)$ are given by projections of the principal components onto the observed ${\bf S}(t) - \langle {\bf S}\rangle$.  

Perhaps surprisingly, variations of a {\em single} principal component account for 99.6\%  and 99.7\% of the observed variability for 2M2139 and SIMP0136 respectively, with second components contributing at the 0.1\% and 0.4\% levels.  In Figure  Fig.~\ref{FigJackiePCAcomps}  the mean spectrum and first two principal components ($\langle {\bf S}\rangle$,$ {\bf E}_0$, and $ {\bf E}_1$), as well as the time-variability of the principal components ($c_i(t)$) are shown for both targets.  In both cases, variations are given by ${\bf S}(t) \approx \langle {\bf S}\rangle + c_0(t){\bf E}_0$, where only a single principal component is required to account for most of the observed variability.  This implies that only {\em two} dominant spectra contribute to the observed variations (e.g., take {\bf E}$_0$ to represent the difference between two types of time-independent spectra ${\bf S}_2-{\bf S}_1$), with the appearance of one "surface type" completely correlated with disappearance of the other.

Thus, our major conclusion is that {\em only two types} of "surface" patches (e.g. cloudy and clear, or thick and thin clouds) are required to explain the observations in both of these sources.  This finding validates a simple light curve model, applied below, in which the photosphere is describe by a linear combination of two 1D model atmospheres differing in cloud thickness and/or temperature.
}

\subsection{Light Curve Analysis}
\label{Mapping}
Next we search for the simplest physically plausible surface brightness model that explains the observed light curve. 
We model the surface brightness distributions using the self-developed genetic algorithm-optimized mapping routine {\em Stratos} (described in Appendix~\ref{Stratos} in detail). The only assumption of the model is that it describes surface features as elliptical spots with their major axes parallel to the rotational direction, {\changes an assumption that is motivated by the common outcome of simulations of hydrodynamical turbulent flows in shallow water approximation \citep[e.g.][]{ChoPolvani1996}. (We note here that our two targets, just like all Solar System giant planets, will be rotationally dominated with a Rossby number R$<<1$, \citealt[][]{ShowmanKaspi2012}).} The input parameters of the model are the number of spots (i) and the number of different surface types allowed (2 in our case, as given by the PCA); the optimized parameters are the inclination and limb darkening of the BD, the surface brightness of each surface type, as well as five values for each spot (longitude, latitude, aspect ratio, area, and surface type). Possible solutions are ranked and optimized on the basis of their fitness, which we define as the reduced chi square difference between the observed and predicted
light curves.

Interestingly, for both 2M2139 and SIMP0136 we find that models with {\em at least} three spots are required to explain the structured light curves: Models with only one or two elliptical spots failed to reproduce the observed light curves. In Fig.~\ref{FigBestModel} we show the best-fit model, a non-unique, but representative solution for 2M2139. Additional models with somewhat different surface distributions are shown in Fig.~\ref{FigDiversity}. Although the solutions are somewhat degenerate (as discussed in Section~\ref{Stratos}) the best solutions for 2M2139 all agree in the following: 1) the overall longitudinal spot covering fraction distribution is similar (between 20\% and 30\%); 2) at least three spots are required;  3) the surface brightness of the spots typically differ by a factor of two to three (either brighter {\em or} fainter) from the surface, corresponding to approximately ~$\simeq$300~K difference in brightness temperature; 4) the size of the largest spot extends about 60$^\circ$ in diameter. We note that more complex solutions with a higher number of spots are also possible, but these replace the larger spots with groups of smaller spots, without changing the general properties identified above.

An important property of all solutions is the presence of very extended spots or spot groups {\changes in the photosphere}, which raises the question whether such large structures can exist in the fast-rotating, warm brown dwarfs. As for Jupiter the gradient of the Coriolis force on our rotating targets is expected to break the atmospheric circulation into parallel jet systems (belts), which will limit the maximum size of continuous atmospheric structures. Here, we use the Rhines scale \citep[e.g.][]{Showman2010} to estimate the relative number of jet systems between our fast-rotating and slowly rotating sources: $N_{jet}\simeq \left(\frac{2 \Omega \times a }{ U} \right)^{1/2}$, where $\Omega$ is the angular velocity, $a$ is the brown dwarf radius, and $U$ is the wind speed. If the maximum size of a feature ($s_i$) is limited by the jet width, the relative maximum spot sizes for two sources will be given by:

\begin{equation}
\label{RelativeSpotSizes}
\frac{s_1}{s_2} = \frac{ \pi / N_{jet,1}}{ \pi / N_{jet,2} } = \left( \frac{\Omega_1 a_1 U_2}{\Omega_2 a_2 U_1} \right)^{1/2}.
\end{equation}

Further, for both sources $a_1$ and $a_2$ should closely approach 1~$R_{Jup}$. If we assume that the wind speeds ($U_1$ and $U_2$ ) are similar in these two T2 dwarfs, the relative spot sizes in Eq~\ref{RelativeSpotSizes} will be simply approximated by $s_1/s_2=(P_2/P_1)^2$, arguing for $\sim$3.2$\times$ larger {\em maximum} spot surfaces on 2M2139 than on SIMP0136, in line with the larger amplitudes observed.

{\changes 
Based on the above considerations we can also estimate the physical spot sizes.
Although wind speeds are not known for brown dwarfs, in the following we explore the possible range. In estimating wind speeds only a few reference points are available: while the highest wind speeds ($U\simeq$2 km/s) yet have been observed in the upper atmospheres of heavily irradiated hot jupiters \citep[e.g.][]{Snellen2010}, much lower wind speeds are typical for the cooler and only weakly irradiated atmospheres of solar system planets (typically 40$-$100m/s, but up to $\simeq$300 m/s in Saturn).  For a discussion on how brown dwarf circulation may fit in this picture we refer the reader to \citet[][]{ShowmanKaspi2012}. 
We will now assume two bracketing case: U=100 m/s (jupiter-like) and U=1,000 m/s (hot jupiter-like). 
Our simple approximation with the low and high wind speeds would suggest $N_{jet}\simeq17$ and $N_{jet}\simeq5$, respectively, for the slowly-rotating 2M2139 and $N_{jet}\simeq32$ and $N_{jet}\simeq10$, respectively, for the rapidly rotating SIMP0136. These values correspond to maximum latitudinal spot diameters of $\simeq10^\circ$ and $\simeq36^\circ$ for 2M2139, and $\simeq6^\circ$ and $\simeq18^\circ$ for SIMP0136. 
Thus, slower, jupiter-like wind speeds would lead in small maximum features sizes (6--10$^\circ$), while high speeds would allow larger features (18--36$^\circ$). While we can not accurately determine the wind speeds in
our targets, the fact that these sources show large-amplitude variations emerging from large regions across their
photospheres argues for wind speeds that are higher than those typical to Jupiter. }


The simple predictions described above are consistent with the much larger variation seen in the slowly rotating 2M2139 than in its faster-rotating sibling and the predicted maximum spot sizes are similar to the size of the largest spots predicted by the best-fit light curve models of 2M2139 (Fig.~\ref{FigSpectraLC}). With more data on varying brown dwarfs a realistic treatment of the atmospheric circulation will be possible, replacing the simple argument introduced above.
                        
\subsection{Atmospheric Model Comparisons}
\label{ModelComparison}

\begin{figure}
\begin{center}
\includegraphics[scale=0.7]{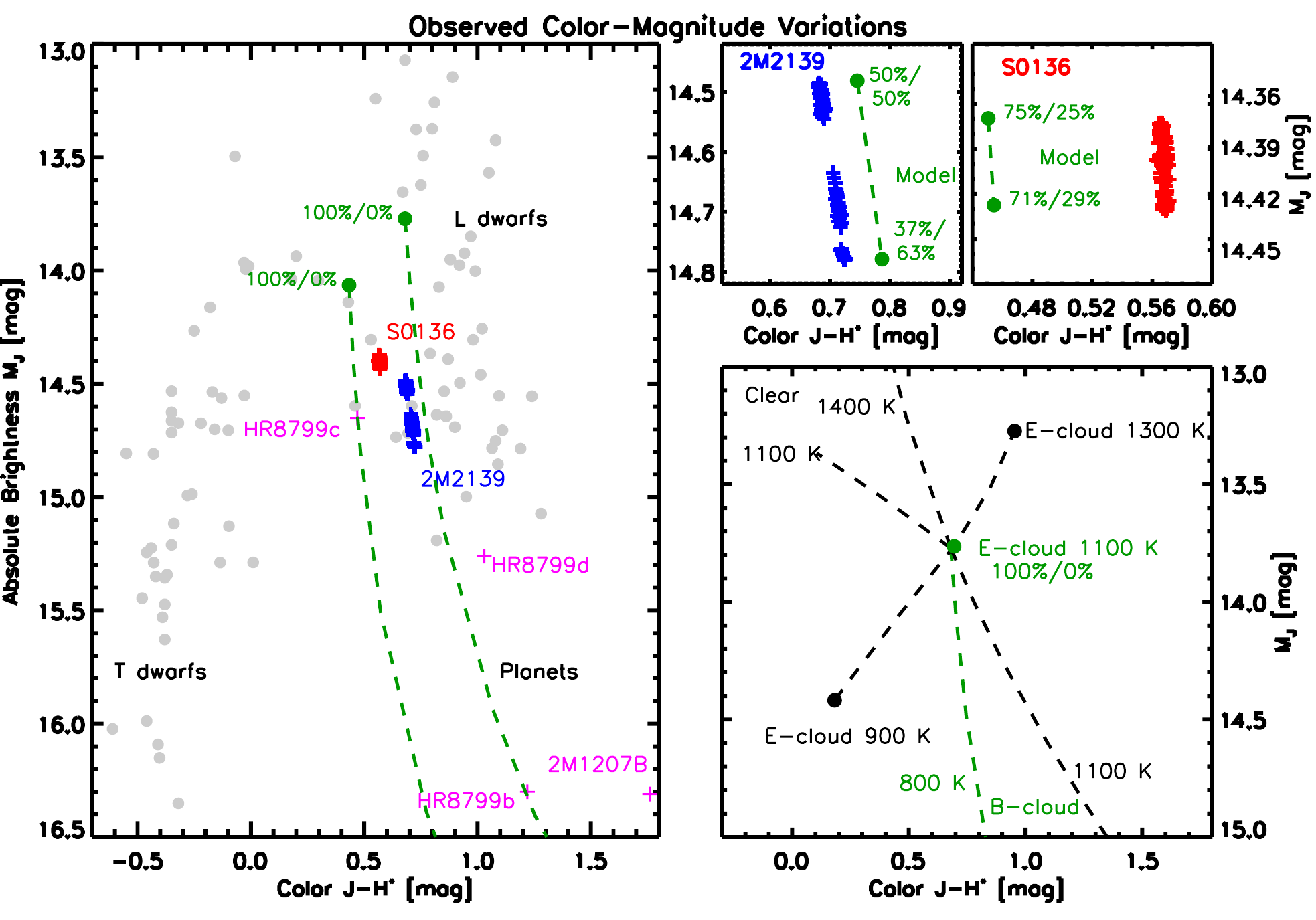}
\caption{The near-infrared color-magnitude variations of our targets (blue and red) in the context of warmer cloudy L-dwarfs and cooler cloud-free T-dwarfs (in gray from \citealt[][]{DupuyLiu2012}). Brightness variations in 2M2139 and SIMP0136 occur without strong color changes (top right and middle right panels). Lower right: Changes predicted by varying single model parameters (black dashed lines) are inconsistent with the observations. The green dashed lines show that simultaneously changing cloud structure (thin to thick) and temperature provides a perfect match. The direction of the modeled changes in 2M2139 and SIMP0136 (green dashed line) is compatible with the poorly understood underluminosity of several directly imaged giant planets (shown in magenta). The percentages show the covering fraction of thin and thick clouds, respectively.\label{FigCMD}}
\end{center}
\end{figure}

Even the most capable state-of-the-art ultracool model atmospheres provide only imperfect fits to the fine structure of brown dwarf spectra; therefore, the full interpretation of our very high signal-to-noise spectral series is constrained by the fact that no existing model can accurately describe atmospheres at sub-percent accuracy. Nevertheless, the existing models can be used to explore the types of changes required to account for the observed color/magnitude variations.
We proceed by identifying the best-fit atmosphere models for our two targets with the assumption that these models would well describe the
dominant surface type on the targets. Then we explore what secondary surface types have to be added to account for the observed color-magnitude  variations by adding a second model atmosphere. 

We base our analysis on the state-of-the-art radiative-convective model atmospheres in and out of chemical equilibrium described in \citet[][]{Burrows2006} and \citet[][]{Madhu2011}, but also used an independent set of models by \citep[][]{Allard2011} to verify that our conclusions are model-independent.
The Burrows models include different empirical descriptions of the vertical structures of clouds of different condensates \citep[][]{Burrows2006}. For each condensate the vertical particle distribution is approximated by a combination of a flat cloud shape function and exponential fall-offs at the high- and low-pressure ends. The cloud altitudes are defined by the intersects of the temperature-pressure profile and the condensation curves. The tested models included cloud shape functions with constant vertical distribution of particles above the cloud base (B-clouds, qualitatively consistent with the DUSTY models described in \citealt[][]{Allard2001}), and different parameterizations of the generic cloud shape function (A, C, D, E in \citealt[][]{Burrows2006}). Of particular interest is cloud type E, a generic cloud with very steep exponential fall-offs corresponding to thin clouds thought to be typical to clouds composed of a single refractory condensate's large  grains. 

\begin{figure}
\includegraphics[scale=0.4]{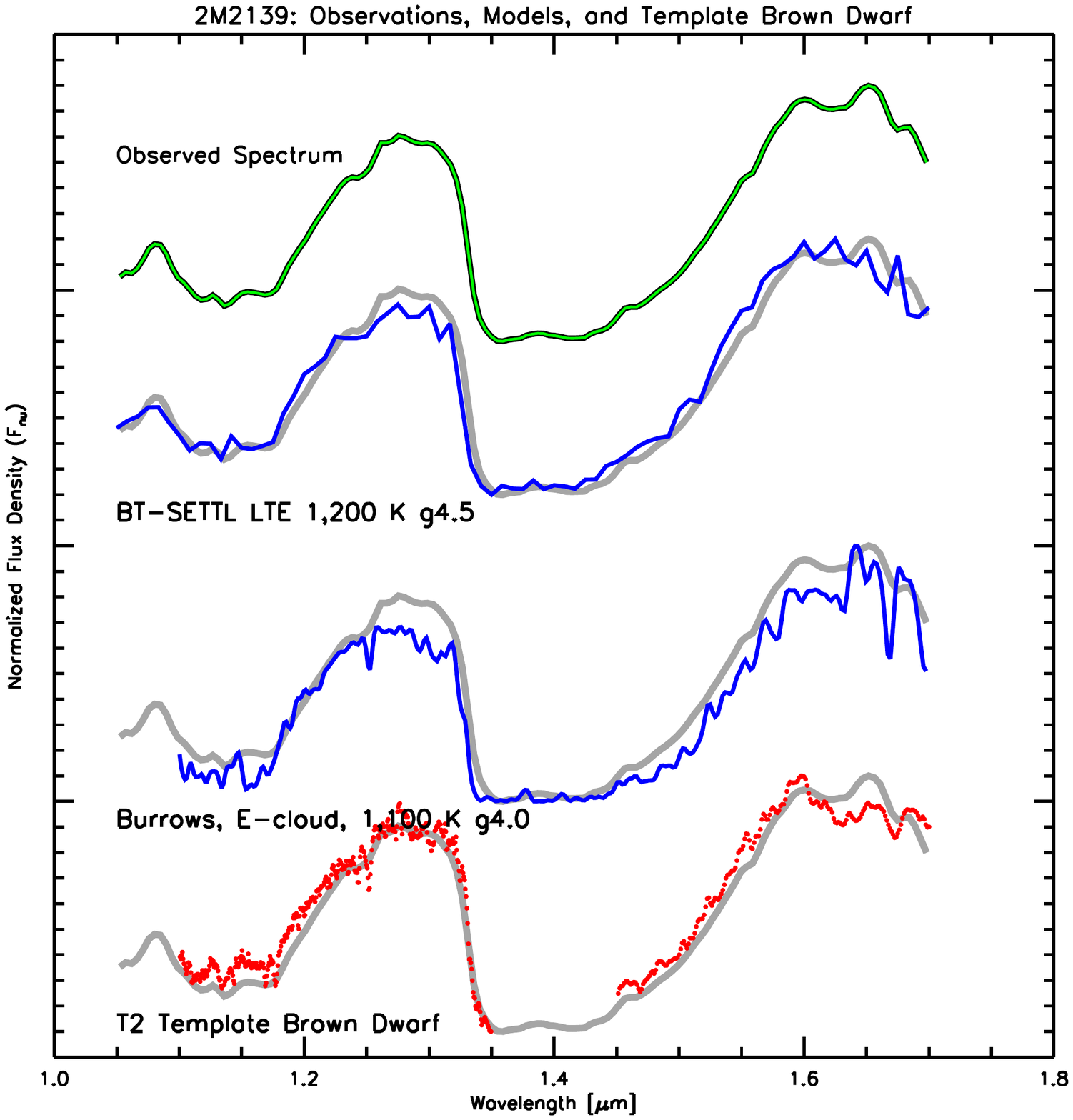}
\includegraphics[scale=0.4]{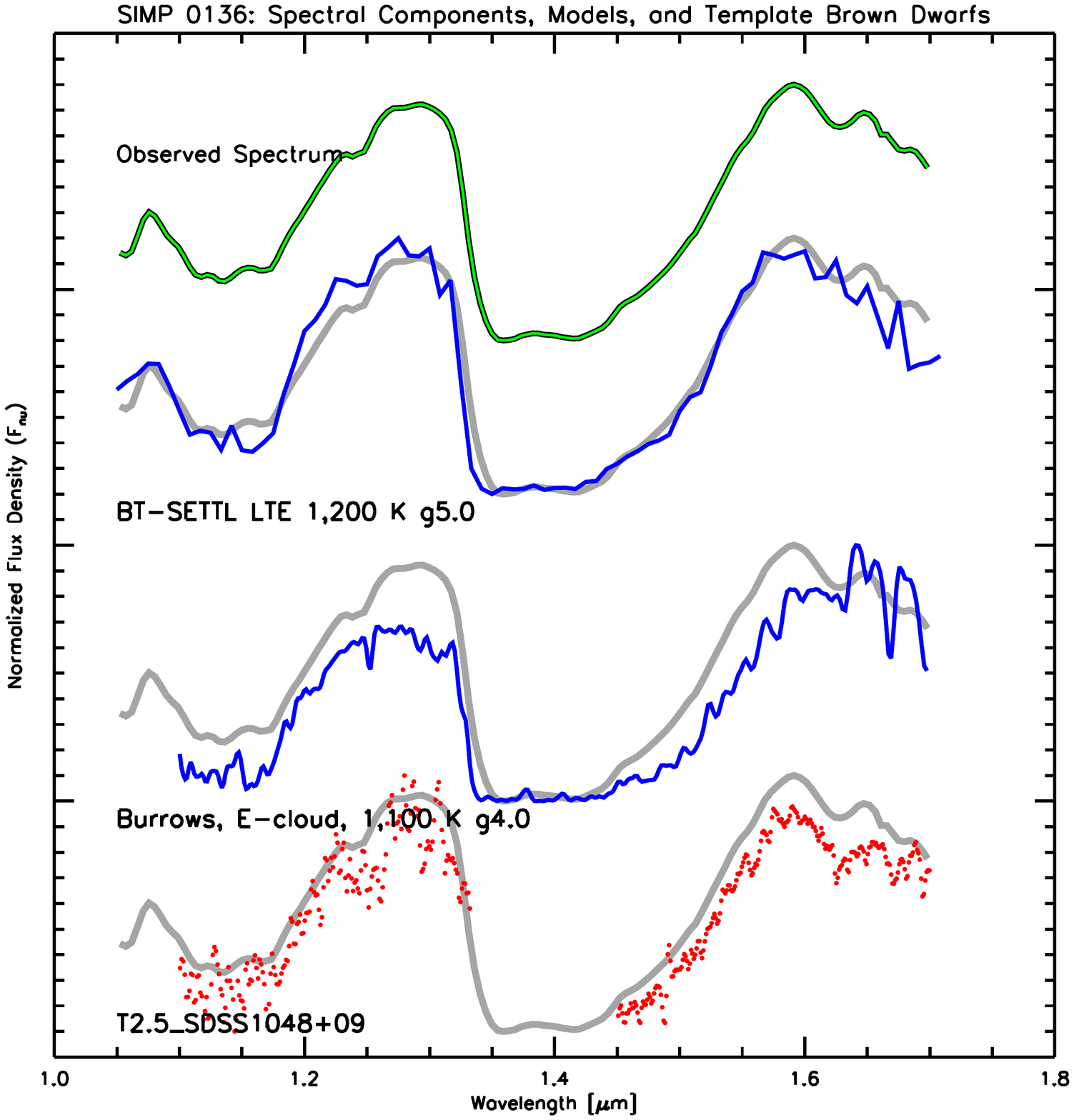}
\caption{Comparison of the observed peak spectra (green), normalized theoretical model atmospheres (blue) and field brown dwarf spectral templates (red). The top model curves are from \citet[][]{Allard2011}, while the lower models are based on \citet[][]{Burrows2006}. \label{FigSpectra}}
\end{figure}

The spectra we explored ranged in temperature from 600~K to 1,800~K with steps of 100~K, included solar and 10$\times$ sub-solar metallicities, and log~g=4.0 and 5.0 for the \citealt[][]{Burrows2006} models and log~g=4.0, 4.5, and 5.0 for the \citealt[][]{Allard2001} models.
Fig.~\ref{FigSpectra} shows our two targets and the best-fit spectra and templates. The upper models are from the BT-SETTL series from \citet[][]{Allard2011}, which we plot for comparison, while the lower curves are model calculations based on models described in \citet[][]{Madhu2011}. The field brown dwarf spectral templates are from \citet[][]{Leggett2000,Chiu2006}. We find
that for 2M2139 the best matching spectral template is a T2 type template, while for SIMP0136 a T2.5 provides a good match.


Both sources can be fit  well with a BT-SETTL model in local thermodynamical equilibrium and an effective temperature of 1,200~K, but fitting the spectrum of 2M2139 {\changes requires a lower surface gravity (log g=4.5) than SIMP0136, which is better matched by a higher surface gravity model (log g=5.0). Fits with the Burrows models suggest slightly lower temperatures. Given the coarser spacing of the surface gravity grid of the Burrows models we used log g=4.0 for 2M2139 and log g=5.0 for SIMP0136. Although the lower surface gravity fits better our spectra, due to its limited wavelength coverage we take these log~g values as comparative but do not argue that they represent
accurate characterization of the surface gravity (see also \citealt[e.g.][]{Cushing2008} for the difficulty of determining precise atmospheric parameters from single-band spectroscopy).}
The modal grain sizes are an adjustable parameter in the Burrows models and models with large grains provided the best match for 2M2139.
We note that this model comparison is based on {\em peak-to-valley normalized} spectra, i.e. focuses on the spectral shape rather than the absolute J, H brightnesses -- we do so because our sources show strong J, H-band variations and only weak variations in the spectral shape.

After establishing the best-fit starting model for the two sources, we explore what secondary surface type is required to explain the observed color-magnitude variations. Our tool for this step is the near-infrared color-magnitude diagram shown in Fig.~\ref{FigCMD}. Here we plotted a full sequence of L/T dwarfs using the parallax and near-infrared photometric database by \citet[][]{DupuyLiu2012}. Fully cloudy L-type brown dwarfs typically appear bright and red, while the dust-free atmospheres of T-type brown dwarfs are blue and faint.  Transition objects are seen to show brighter J-band magnitudes with later spectral types before an eventual turn to the clear T-dwarf sequence \citep[][]{Dahn2002,Tinney2003,Vrba2004}.

Next, we tried to vary each model parameter separately to assess its effect with the observed variations. We found that {\em no} combination of models with a difference only in a single parameter (temperature, cloud scale height, presence or absence of cloud layer) can introduce the observed changes (see dashed black lines lower right panel in Fig.~\ref{FigCMD}): the tracks along which the source would vary on the CMD if such a secondary surface type would be added is clearly inconsistent with the actual variations. The same conclusion was reached by \citet[][]{Artigau2009} for
SIMP0136 based on comparison of models by \citet[][]{Allard2003} and \citet[][]{Tsuji2005} to their ground-based $\Delta$J/$\Delta$K observations and by \citet[][]{Radigan2012} for 2M2139 through comparison to models by \citet[][]{SaumonMarley2008}.

We next explored {\em correlated} changes in parameter pairs. We found that correlated changes in cloud scale height and temperature are required (green dashed lines in Fig.~\ref{FigCMD}). By allowing these two model parameters to vary together we find that thin clouds in combination with large patches of cold and thick clouds (i.e. T$_{eff}=1,100$~K models with E-type clouds and 800~K B-type clouds, green dashed lines) can explain well the observed color-magnitude variations (blue and red crosses). 
                                                                      
This solution requires about 300~K temperature difference between the spots and the surface, fully consistent with the factor of three surface brightness change predicted by our light curve shape model. Our models also predict the relative surface covering fraction for the thin and thick clouds (given as percentages in Fig.~\ref{FigCMD}), which are also consistent with the surface model shown in Fig.~\ref{FigBestModel}.

We note that qualitatively similar results were obtained for 2M2139 using a ground-based photometric dataset and models of \citet[][]{SaumonMarley2008} by \citet[][]{Radigan2012}.

\begin{figure}
\includegraphics[angle=0,scale=1.00]{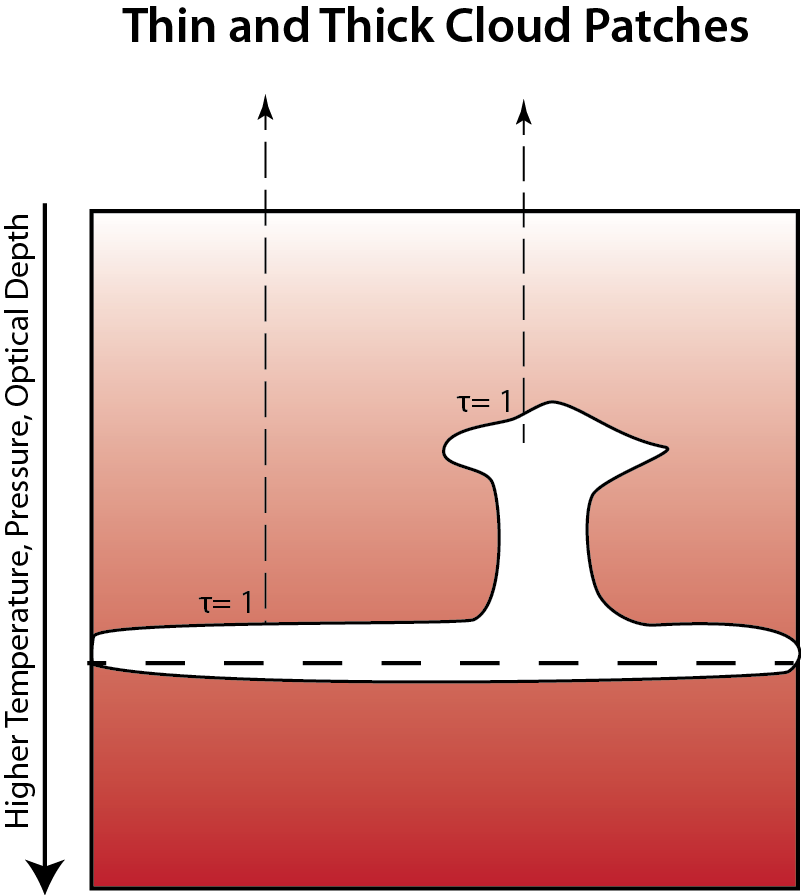}
\caption{Sketch illustrating a possible cloud structure consistent with the observations,
which argue for large-scale variations in dust cloud scale height, correlated
with a change in temperature. Higher clouds will limit the observed column to the cooler upper 
atmospheres, explaining the correlated changes in temperature and cloud scale height.  More
complex configurations, such as multi-layer clouds, are also possible. \label{FigSketch}}
\end{figure}

\section{Discussion}

\subsection{A Single Spot Type}

Because of the richness of possible condensates in brown dwarf atmospheres one may expect a complex mix of cloud properties, composition,  and {\changes cloud} scale height to be represented even within a single brown dwarf atmosphere. The observational fingerprint of such surface complexity in a rotating brown dwarf would be multi-component, complex changes in the spectra, brightness, and color. In stark contrast, the two sources observed show a single distinct type of simple change. This is reflected, for example, in the single-direction track in the color-magnitude diagram (Fig.~\ref{FigCMD}). The PCA analysis (see Section~\ref{PCA}) reveals that all the observed variance in the two sources can be reproduced with the combination of {\em only two} different spectra. This surprising result shows that all major features in the visible photosphere, although distributed across both hemispheres of each target, share the same spectra (i.e. same deviation from the mean spectrum).

 We note that based on spectral fits from composite near-infrared spectra \citet{Burgasser2010} has identified 2M2139  as a strong candidate for being an unresolved binary brown dwarf. These authors argue that 2M2139 could be matched better, although imperfectly, by a blend of an L7-L9.5 and a T3.5--T4.0 binary than by any single template brown dwarf. It is tempting to consider the possibility that the fit by two blended spectra in this case is not the sign of an unresolved binary, but instead emerges from the blend of two different surface types on the source (as also proposed by \citealt[][]{Khandrika2013} and \citealt[][]{Radigan2012})--- one with a spectrum of an L8.5$\pm$0.7 dwarf, the other with a spectrum of a T3.5$\pm$1.0 dwarf. However, this explanation is unlikely
to be correct. A composite spectrum of two such dwarfs with time-varying weights due to the rotation would introduce
 J vs. J--H color variations in the direction of the L--T  transition, very different to what we observed (see~Fig.~\ref{FigCMD}). 

The fact that flux density in the observed molecular bands -- with the exception of the less-changing water band -- varies together and at the same rate as the continuum shows that the opacity variations in the targets {\em cannot} originate from changes in the abundances of the common gas-phase absorbers, such as methane.
Based on comparison to atmosphere models we argue that these spots are patches of very thick clouds. The fact that any two such cloud patch would share the same spectra, different from the dominant spectrum of the sources, argues for the presence of a {\em single mechanism} to form these thick cloud patches. Possibilities, as explored below, include circulation and large-scale vertical mixing.

\subsection{Complex Surfaces}
With our surface mapping tools (see Section~\ref{Mapping}) we find that both sources have relatively complex surface brightness distributions: assuming
elliptical spots, no one- or two-spot model can reproduce the structure of the light curves. While our model is not able to deduce the {\changes precise} appearance of the {\changes photosphere}, it provides useful insights into the complexity and overall distribution of the thick cloud patches observed. Specifically, the lightcurve and our 
models reveal that a large fraction of the surface of both targets is covered by the cloud patches. These patches may be very large single structures (corresponding to our simplest model) or they may be super-structures,  consisting of dozens or even hundreds of smaller cloud patches with the same surface covering fractions. Whether single or complex structures, however, any mechanism that explains the spectral appearance and high cloud scale height, needs to account for the concentration of these patches on the surface of the targeted brown dwarfs.

{\changes   We note here that the lightcurve modeling is only sensitive to the varying surface components, because homogeneously 
distributed features will not introduce brightness variations. Thus, it is likely that the large patches deduced from the light curve are
not the only features present in the photosphere. Reinforcing this possibility is the fact
that color-magnitude modeling based on model atmospheres (Section~\ref{ModelComparison}) suggests similar level of asymmetry but on
top of a more symmetric component.  According to the best-fit model atmosphere combination the fraction of surface covered by thick clouds 
varies from about 50\% to 63\% on the visible 2M2139, while it occupies only 25\% to 29\% of the surface of SIMP0136
(see Fig.~\ref{FigCMD}). Thus, the modeling suggests that most of 2M2139's surface is covered by thick clouds with large thin patches (that are brighter and contribute most of the observed emission), while SIMP0136 has an overall thin cloud layer with large patches of thick clouds. Although the covering fractions are different, based on the overall similarity of the spectra and the spectral variations these two sources have very similar thin and thick cloud layers.}

The complex distribution of otherwise similar or identical thick cloud patches also offers opportunity for further exploration of the dynamics of brown dwarf atmospheres. Cloud structures are subject to multiple {\changes dynamical} processes and are likely to evolve on a broad range of timescales. For example, the structure of the light curves may evolve due to differential rotation, an effect that should be observable with high-precision dataset covering sufficient baselines. Other processes, such as thick cloud formation (i.e. rapid increase of the cloud scale height) or the reverse of this process, rain-out of the condensate grains, may also result in changes over relatively short timescales.

\subsection{Thin and Thick Clouds, but No Deep Holes}

A leading hypothesis to explain the dramatic spectral changes observed to occur at the L/T transition invokes a breaking apart of the cloud cover \citep[e.g.][]{AckermanMarley2001,Burgasser2002}  This idea provided motivation for our initial observations, and predicts surfaces consisting of clouds and holes that act as windows into the deeper photosphere.  The most simple picture, and the one that has been explored by models \citep[][]{Burgasser2002,Marley2010} is one where holes in the cloud layer represent pure clearings, 100\% free of condensate opacity.  In contrast to this assumption, our observations of two of the most variable brown dwarfs suggest that dark and bright regions of the photosphere represent thick and thin cloud regions rather than cloudy and cloud-free regions (Section~\ref{ModelComparison}).  This may reflect holes in a thick cloud layer that look down into a thinner cloud layer.    More generally, our observations argue for a more complex picture of cloud heterogeneities than envisioned with simple cloudy and cloud-free models.  A similar conclusion was arrived at by Radigan et al. (2012), who found that JHK photometric monitoring of 2M2139 was inconsistent with the presence of cloud-free regions, based on models of \citet[][]{SaumonMarley2008}.  Thus both photometric observations out to the K band and spectroscopic observations from 1-1.7 microns argue against the existence of cloud-free regions.

We also found that the cloud thickness variations are correlated with changes in the effective temperature of the secondary model. This correlation is not surprising: the higher the dust scale height, the shallower pressures and the lower temperatures are visible to the external observer (see Fig.~\ref{FigSketch}). This correlation, thus, argues for patches of thick clouds towering over the otherwise thinner cloud layer covering most of the hemispheres of the targets. 

\begin{center}   
\begin{figure}
\includegraphics[scale=0.86]{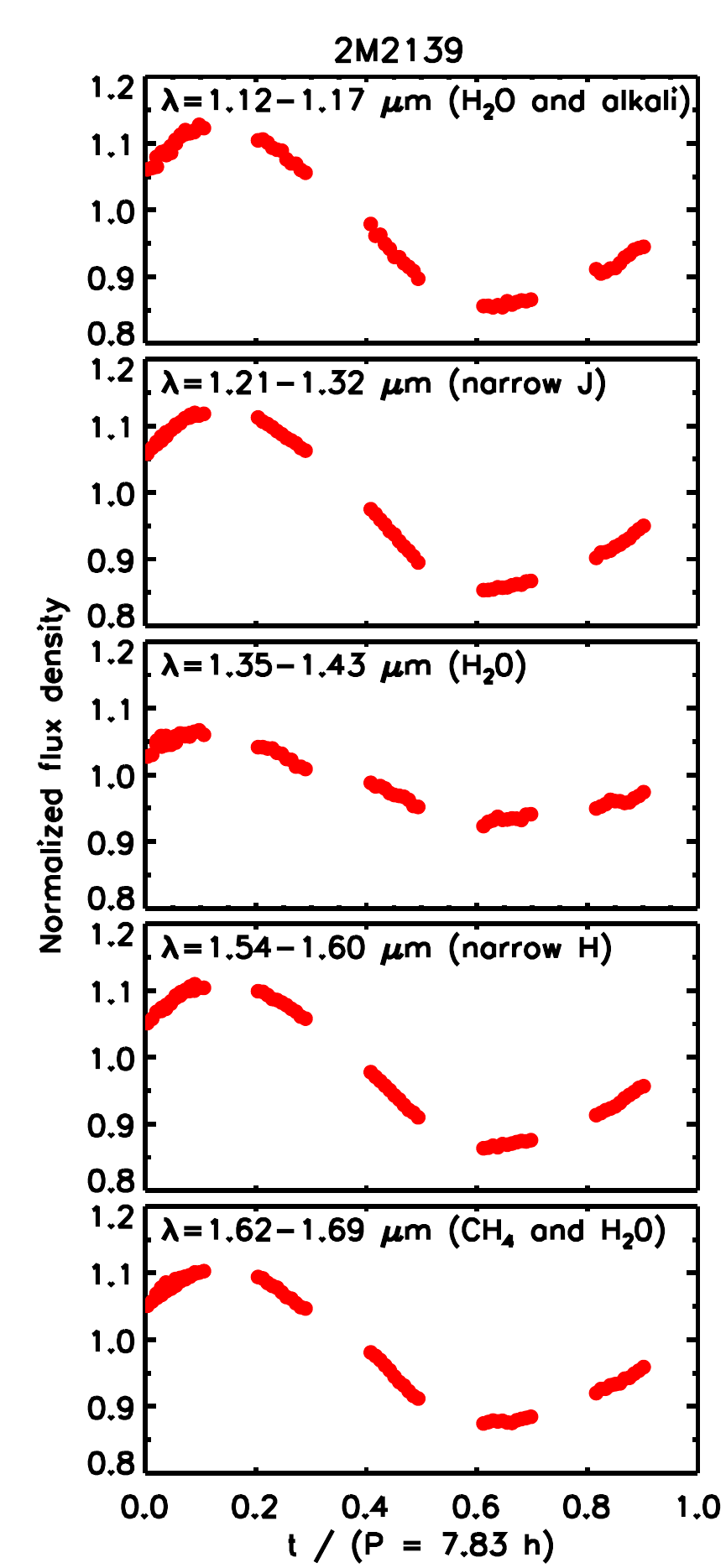}
\includegraphics[scale=0.86]{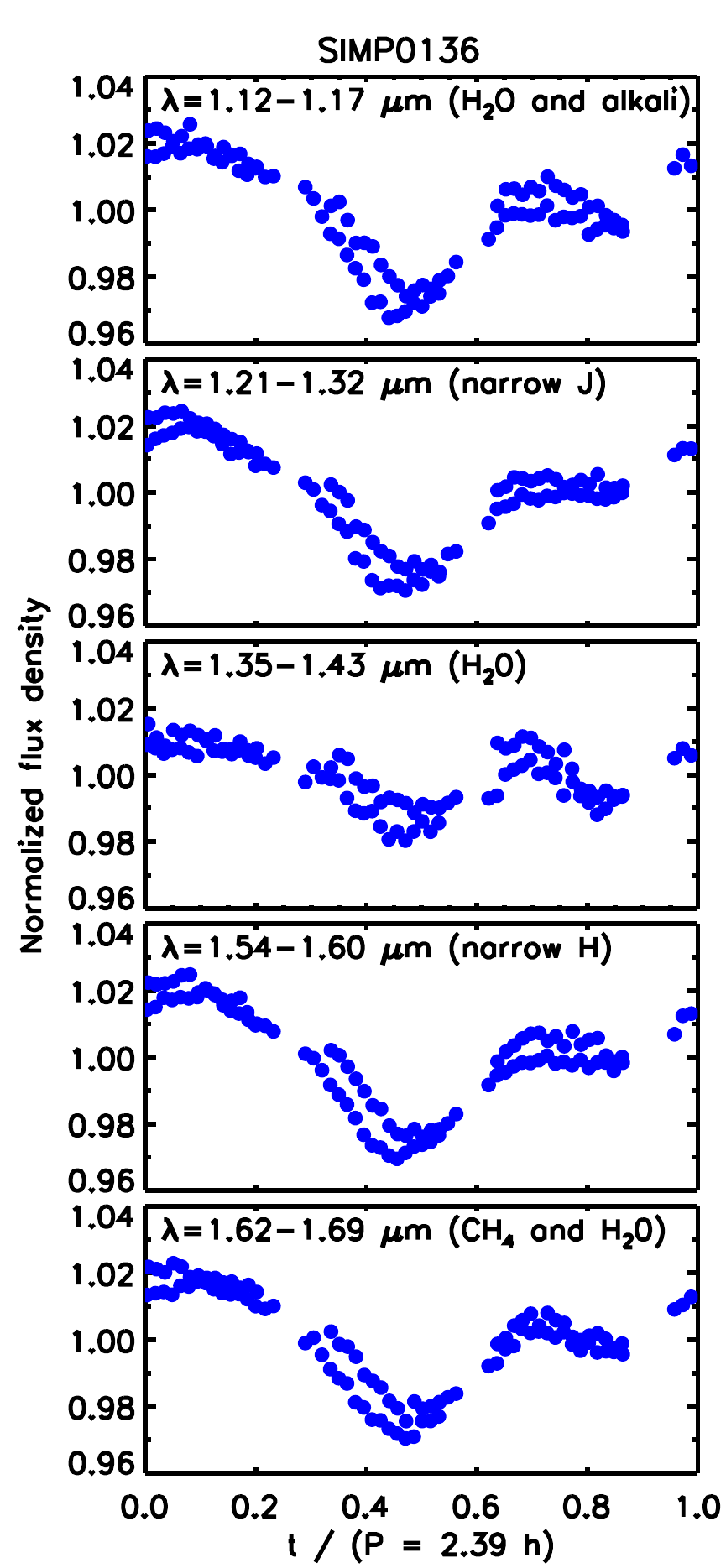}
\caption{Five narrow-band light curves extracted from the spectral series for the two targets show that the LC changes are all
occurring at the same phase. The spectral bands have been selected to probe specific atmospheric depth and match those 
adopted by \citet[][]{Buenzli2012}. In contrast to the no phase shift seen in our two T2 dwarf, the T6.5 dwarf analyzed by \citet[][]{Buenzli2012}
showed a very prominent pressure-dependent phase shift in the same narrow-band light curves, revealing a large scale horizontal-vertical
structure. \label{Nophaseshift}}
\end{figure}
\end{center}

The vertical structure and composition of these thick clouds is an {\changes exciting} question, albeit one our current data does not constrain well. {\changes Recently \citet[][]{Buenzli2012} have shown that by combining data over a broad wavelength range -- where different wavelengths probe different atmospheric depths -- the vertical structures of the clouds can be explored. In their study five atmospheric layers of a T6.5 dwarf's were sampled by obtaining five complete light curves at depths ranging from 0.1 to $\sim$10~bar. 
An important and surprising result of their study was a significant and pressure-dependent phase difference in the atmosphere with
the largest phase shift observed at the deepest level exceeding 180$^\circ$. While a full analysis like that carried out by  \citet[][]{Buenzli2012} is beyond the scope of this paper, we show in Fig.~\ref{Nophaseshift} that the same narrow-band light curves as used by those authors, when extracted from our spectral series, show no significant phase shift.
Thus, while the results of the T6.5 brown dwarf suggest correlated large-scale vertical-horizontal structures, the two early-T dwarfs studied here show similar cloud structures at different layers without phase differences.}
   
\subsection{Apparent Underluminosity due to Thick Clouds}

The remarkable cloud scale height variations in our targets demonstrate how this parameter affects the brightness of ultracool atmospheres. 
Thick clouds persisting at temperatures lower than typical for brown dwarfs have been proposed as one of the processes that may explain the apparent underluminosity of directly imaged giant exoplanets  compared to brown dwarfs with similar spectral morphology (magenta crosses in Fig.~\ref{FigCMD}, see also \citealt[][]{Skemer2012,Barman2011_HR8799,Barman2011_2M1207,Currie2011,Madhu2011,Marley2012}). Several authors propose that such unusually thick clouds would be present in directly imaged exoplanets due to the low surface gravities of these sources, which provides an attractive and self-consistent explanation for the appearance and low occurrence rate of these sources. However, the effects of thick clouds remained difficult to verify, as multiple parameters (metallicity, surface gravity, age, mass, chemistry, cloud structure) vary simultaneously between any two ultracool atmosphere. The novelty of our observations is that they are comparing {\em different cloud structures} within the {\em same atmospheres}, i.e. keeping metallicity, surface gravity, age, mass, {\changes and bulk composition} constant. This allows isolating the effects of cloud structure. {\changes We note that a minor caveat here is that the pressure-temperature profile of atmospheres with large spots may differ from those with only thick {\em or} thin clouds; thus, the thick clouds' impact must be evaluated keeping this possible difference in mind. } 

We can explore the effect the thick clouds would have by extending their surface covering fraction in our model beyond that observed in our targets, i.e. by increasing the surface covering fractions of the thick cloud patches to values approaching 100\%. The resulting green dashed line in Fig.~\ref{FigCMD} shows that increasing thick cloud coverage will produce color-magnitude tracks crossing the positions of the directly imaged planets (magenta crosses). Thus, if the thick cloud patches we observed in the atmospheres of our T2 targets would cover their complete or near-complete atmospheres, their near-infrared brightness and colors would provide a good match to the directly imaged exoplanets. These results lend support to the models in which  the peculiar color and brightness of exoplanets is introduced by thick clouds, but we note that in this work no attempt was made to match the spectra predicted by our
simple models to those observed by exoplanets. While our observations cannot identify the cause for the thick clouds in exoplanets,  we expect that any successful model for the thick clouds in faint and red exoplanet atmospheres will also provide an explanation for the coexistence of thin and thick clouds in L/T transition brown dwarfs.

\subsection{The Path to Next Generation Atmosphere Models}

Our observations provide exceptionally high signal-to-noise spectra that probe spectral variations within the photospheres of brown dwarfs. The accuracy of this dataset is high enough that finding a perfect match ($<1\%$) with existing atmosphere models was not possible, demonstrating the limitations of the existing models. Although the best fit spectra match the mean spectrum to
typically within 5--10\% in the spectral range studied, which are typically considered to be good fits for brown dwarf atmospheres, these differences are comparable to or larger than the amplitude of changes we observe in our spectral series. Therefore, the accuracy of the existing models used in this paper did not allow meaningful modeling of the entire spectral series yet.  Instead, in our modeling procedure we started from atmosphere models that provided best fits to the mean spectra of the targets and then explored their {\em color variations}. Our approach to model a patchy cloud cover with a linear combination of independent 1D models is imperfect and it is likely not physically self-consistent due to the somewhat different pressure-temperature profiles of these models \citep[e.g.][]{Marley2010, Marley2012}. 

It is  worthwhile to briefly explore the limitations of current models and potential pathways to improve them. Arguably, the key limitations are the incomplete molecular opacity databases and the fundamentally one-dimensional nature of most atmosphere models. Expanding and refining the opacity databases relies on the continuation of ongoing laboratory and theoretical efforts and not limited by the observations of ultracool atmospheres. 
In contrast, constraining and further developing two- or three-dimensional models (e.g. \citealt[][]{Freytag2010}) will require more accurate datasets and, in particular, datasets that provide spectrally and spatially resolved information. 
We anticipate that the brown dwarf spectral mapping technique introduced in this paper will lead to a major step in testing and refining physically realistic models of cloud 
formation and cloud structure \citep[e.g.][]{Helling2008a,Helling2008b} and atmospheric dynamics \citep[][]{Freytag2010,Showman2013}

\subsection{Spectral Mapping of Ultracool Atmospheres}

 In this paper we also applied a method, spectral mapping, to a new class of objects, ultracool atmospheres. As a new
method for this field we now briefly discuss its potential and future uses.  Photometric 
phase mapping has been proposed early to reach spatial information beyond the diffraction limit \citep[e.g.][]{Russell1906}. Since then,
different variants of this idea have been used successfully to derive asteroid shapes from photometry \citep[e.g.][]{Kaasalainen2001,Kaasalainen2012}, to map their surface composition from spectral mapping \cite[e.g.][]{Binzel1995}, 
to map starspots via photometry and Doppler imaging \citep[e.g.][]{Budding1977,Vogt1987, Luftinger2010}, and to translate precision Spitzer photometry to one-dimensional and two-dimensional brightness distributions for hot jupiters \citep[e.g.][]{Knutson2007,Cowan2012a,Majeau2012}. Similarly to the hot jupiter studies, in an upcoming study Heinze et al. (ApJ, submitted) use precision Spitzer photometry complemented by ground-based near-infrared lightcurve to explore cloud properties on an early L-dwarf.
Most recently, \cite[][]{Buenzli2012} have found pressure-dependent phase shifts in multi-band {light curves} extracted from HST spectral series and a Spitzer 4.5~$\mu$m light curve of a T6.5 brown dwarf, revealing vertical structure in an ultracool atmosphere for the first time.  

The method used here is a logical next step in these brown dwarf studies, where a large time-resolved spectral set is used to 
to identify the diversity, spatial distribution, and spectra of the key photospheric features. Further similar studies
with space-based instruments on HST and Spitzer, complemented by sensitive ground-based photometric observations,
will allow obtaining data similar to those presented here on brown dwarfs covering a much broader range
of atmospheric parameters. Such a dataset will allow the exploration of the properties of cloud cover with spectral
type, surface gravity, and rotation period, an important step toward establishing a physically consistent picture of
condensate clouds. Next-generation adaptive optics systems 
will also be capable of measuring relatively small photometric variations in directly imaged giant 
exoplanets, allowing comparative studies of brown dwarfs and extrasolar giant planets \citep[][]{KostovApai2012}. 

The changes observed in SIMP0136, first reported in \citet[][]{Artigau2009} and also seen in our Fig.~\ref{FigSpectraLC}, also highlight another exciting question. Cloud covers in some brown dwarfs 
clearly change on very short timescales on very significant levels, offering an opportunity to study 
the atmospheric dynamics of ultracool atmospheres via multi-epoch, multi-timescale, multi-wavelength 
observations. 

Spectral mapping is set to be a powerful new method not only to characterize brown dwarf atmospheres, but
also extrasolar planets. 


\section{Summary}

In summary, we apply spectral mapping for the first time on ultracool atmospheres and show that two L/T transition brown dwarfs have patchy cloud covers with multiple ($>$3) large spots/structures. Analysis of the spectral variations shows that a linear combination of only two types of spectra can explain the variance observed in both sources, demonstrating the presence of a single type of photospheric features in an otherwise homogeneous cloud cover. We find that light curves derived from narrow wavelength sections of the spectra all change in
phase.
The observed variations show that the near-infrared brightness of dusty brown dwarfs can decrease significantly  (3--27\%) with only a modest reddening in the J-H color. These changes and extrapolation from the atmospheric models fitting them closely resemble the properties of red and "underluminous" directly imaged exoplanets, arguing for thick clouds causing the underluminosity of giant planets. Our models with large cloud thickness variations and correlated temperature variations ($\simeq$300~K) explain the observed light curves (amplitudes, color-magnitude changes) as well as the light curve structures. Our findings reinforce models that explain the underluminosity of directly imaged super-jupiters with large scale height dusty atmospheres. 
 
The technique applied here, rotational phase mapping, provides a powerful tool to study the atmospheres of ultracool objects, brown dwarfs and exoplanets. 
Cooler objects may harbor clouds of particles with various composition (NH$_3$, CH$_4$, H$_2$O) and phase (solid or liquid), the presence of which can be inferred with this technique. Its full potential, however, can be achieved with a telescope like James Webb Space Telescope, which will couple high sensitivity and broad wavelength coverage with high contrast, enabling spectrally and spatially resolved mapping of directly imaged exoplanets.

\acknowledgments
We acknowledge the anonymous referee, whose comments improved the manuscript.
Support for Program number GO-12314 was provided by NASA through a grant from the Space Telescope Science Institute, which is operated by the Association of Universities for Research in Astronomy, Incorporated, under NASA contract NAS5-26555.
The work of JR and RJ was supported by grants from the Natural Sciences and Engineering Research Council of Canada.

We are grateful to the dedicated staff at the Space Telescope Science Institute for their outstanding support of the observations and the instrumentation. 

Calibrated data and reference files used in this work are available indefinitely at the B. A. Mikulski Archive for Space Telescopes (http://archive.stsci.edu). We acknowledge an STScI DirectorÕs Discretionary Grant that helped the start of this project.

{\it Facilities:} \facility{HST (WFC3)}.


\appendix
\section{Surface Mapping: The Stratos Package}
\label{Stratos}

Translating a light curve into the surface map of a rotating sphere is an under-constrained deconvolution problem. Although due to the nature of the problem not all information can be retrieved, equipped with a few priors about the physically plausible geometries and a very high signal-to-noise data it is possible to derive meaningful maps with robust characteristics.

To model our data we developed a new IDL package ({\em Stratos}). In the following we briefly describe the principles and organization of this package and then discuss the solutions and degeneracies in the derived parameters.
In {\em Stratos} we apply an optimized forward-modeling procedure to identify the best-fitting two-dimensional map for each spectral series given a small set of a priori assumptions. 

We start by using principal components analysis to identify $i$, the number of independent spectral components required to explain at least 96\% of the observed spectral variance (see Sect.~\ref{PCA} for both of our sources $i=2$). This determines the number of the types of surface features we include in our model, including the ambient spectrum. Our actual surface model consists of a sphere with $j$ ellipses {\changes in its photosphere}. The surface brightness level of each of these ellipses is fixed to one of the i distinct levels. Thus, the basic free parameters of the model are the latitude, longitude, axes ratio, and area of each ellipse, i.e. four parameters per ellipse. The relative brightness of the {\em i} different surface types are also free parameters (in our case only a single parameter). In addition to these parameters we add the inclination of the targetÕs spin axis and the limb darkening as free parameters. We express limb darkening in the commonly used form as $I(\phi)=I_0 \times (1-c\times(1-cos(\phi))$, where $\phi$ is the angle between the line of sight and the observed surface element and $c$ is the limb darkening coefficient fitted. We allowed c to vary between 0 and 0.8, the former representing no limb darkening and latter upper bound corresponding to the strongest limb darkening
predicted in near-infrared wavelengths for low-mass stars \citep[][]{Claret2011}.

We optimize the above model using a genetic algorithm, a commonly applied heuristic optimization algorithm (see \citealt[e.g.][]{Charbonneau1995}). We define the fitness of each solution by the sum of the squared differences between the predicted and observed spectral variations. Multiple parallel optimization runs were executed on a 12-core Intel Xeon-based Mac Pro; typical runtimes are about two days per target.

\begin{figure}
\begin{center}
\includegraphics[angle=0,scale=0.45]{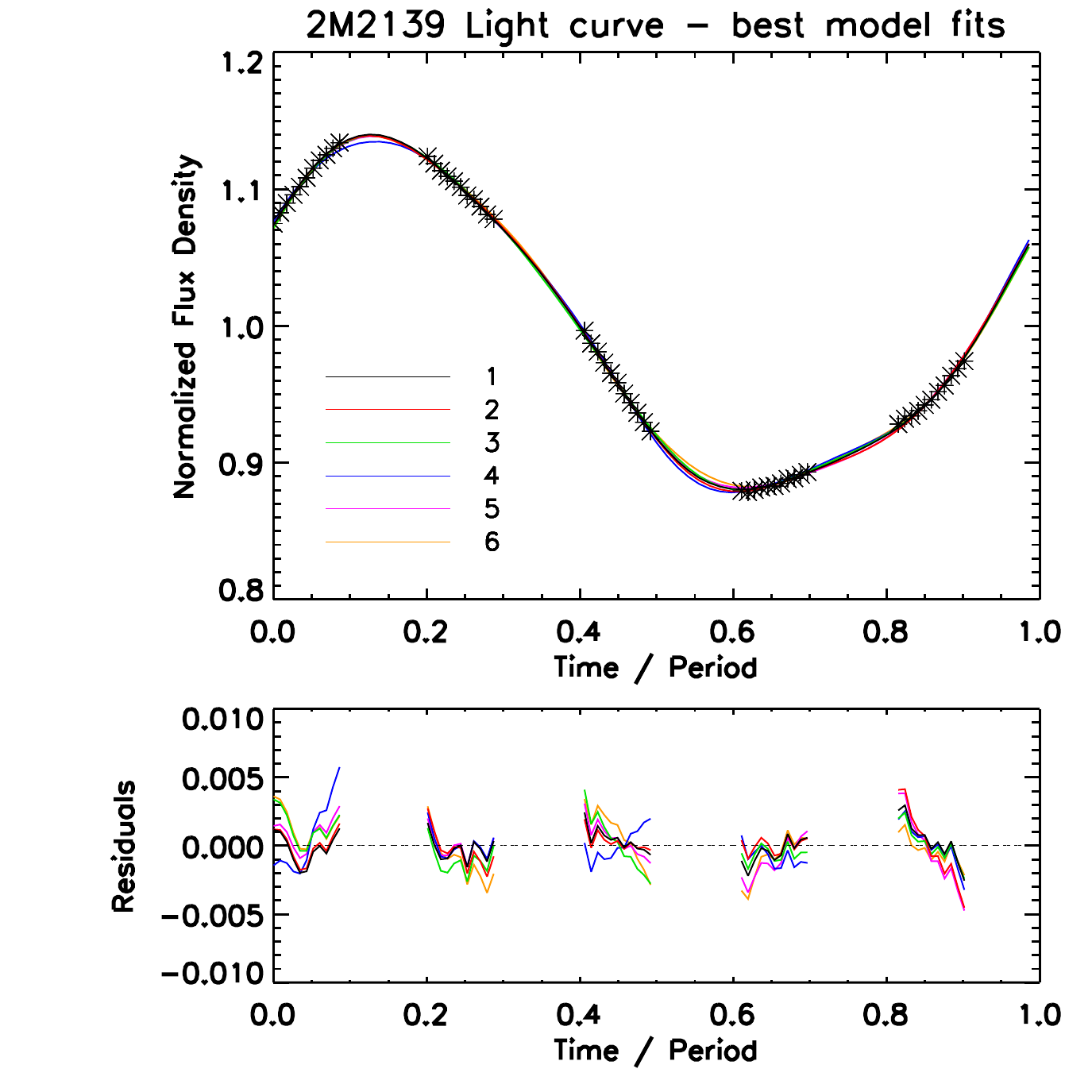}
\includegraphics[angle=0,scale=0.55]{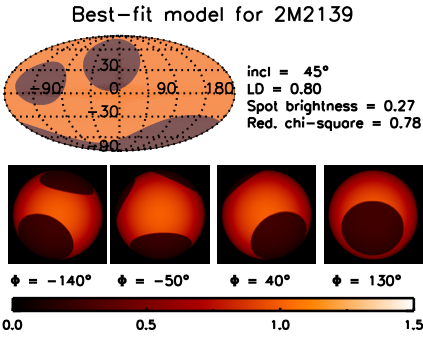}
\caption{{\em Left:} The Stratos package provides multiple, broadly similar models that can reproduce well the observed light curve of 2M2139. All models require at least three distinct spots. Models are numbered by increasing chi square and 
shown in Fig.~\ref{FigDiversity}. {\em Right:} Example best-fit model for 2M2139. Although some parameters are degenerate,
all models require multiple large spots distributed across the surface of the source in a broadly similar pattern.  \label{FigBestModel}}
\end{center}
\end{figure}

\begin{figure}
\begin{center}
\includegraphics[angle=0,scale=0.55]{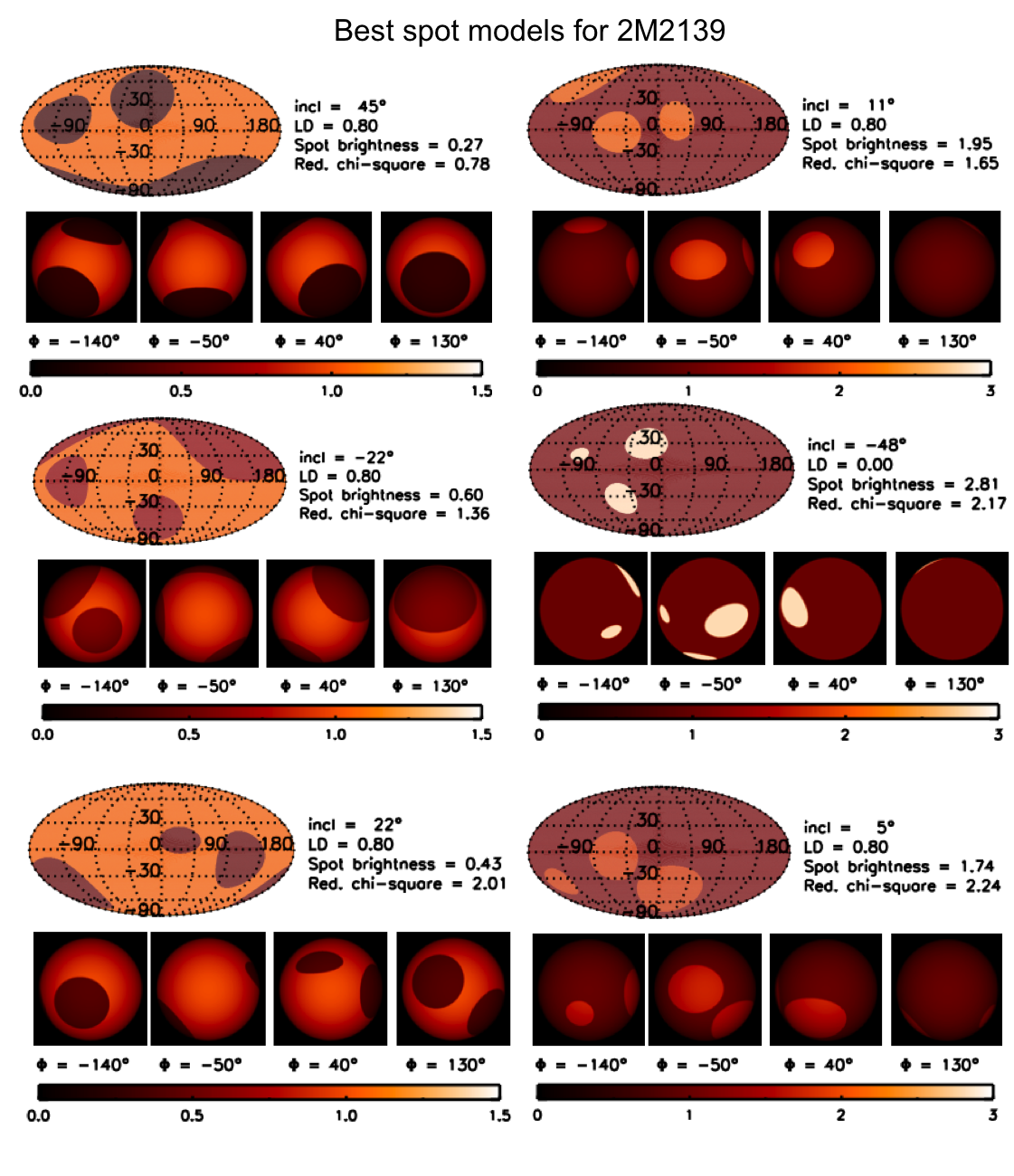}
\caption{Although light curve decomposition yields degenerate solutions, several key properties of
the solutions are similar. No solution with less than three spots can fit the data; the relative spot sizes and total covering
fractions of the models are very similar. When the inclinations are considered the spot distributions are also similar.
While no unique solution exists, the modeling provides an insight into the similarities of the simplest best-fitting models. 
\label{FigDiversity}}
\end{center}
\end{figure}

The maps derived with phase mapping and Stratos have three natural limitations. First, our data is sensitive to variations in the surface brightness distributions and insensitive to time-invariant features. Second, the observations can only probe the visible fraction of the {\changes photosphere}: for a rotation axis inclined with respect to the plane of the sky part of the {\changes photosphere} will not be visible at any rotation phase. Third, a variation in the latitude of any feature only slightly changes the light curve.

The above limitations lead to solutions that are degenerate in some parameters, while robust in others. In the following we discuss the similarities and degeneracies in the best-fit solutions.


During a typical fitting procedure Stratos evaluates about $\simeq10^5$ different solutions. Figure~\ref{FigDiversity} shows the light curves for the six best-fit models for 2M2139, the target with the highest signal-to-noise levels. Note, that the light curves predicted by the six models differ the most in the inter-orbit gaps, where HST could not obtain data. These models provide excellent fits and the amplitude of the residuals are less than 0.5\% (Figure~\ref{FigBestModel}).

Figure~\ref{FigDiversity} provides an overview of the six best-fit surface maps. For the first glance these solutions may appear different, but even cursory inspection reveals that all solutions share several key characteristics. First, three spots provide excellent fits to the observations, whereas no one- or two-spot solution was acceptable. Second, the longitudinal distributions of the visible surface area occupied by lighter and darker features are very similar. Third, the projected sizes and relative positions of the two larger spots are also similar when considering the inclination of the model.
Clearly, there are also several parameter pairs that are degenerate and not tightly constrained: the inclination and the limb darkening and, to a lesser degree, the spot size and the spot surface brightness.

{\em Bright spot/dark spot degeneracy:} Our surface modeling procedure can model the light curves equally good with large bright spots on darker surface or with large darker spots on a lighter surface. These solutions -- when the inclination and the latitudinal integration is considered -- appear to be inverse of each other and, ultimately, lead to the same longitudinal one-dimensional surface brightness distribution.
At the level of accuracy of our measurements these two families of solutions are degenerate, although they may be distinguishable in the future with more precise datasets and a shape model for the features.
In the context of our atmospheric modeling the darker regions can be interpreted as covered by thick clouds, while the brighter regions are covered by thinner clouds with a higher-pressure upper boundary (i.e. warmer). Whether the photosphere should be interpreted as
a thin cloud layer with towering thick clouds or a thick cloud layer with depressions or cavities (but not deep holes) depends on the covering
fraction of the two surface types. The atmospheric modeling (see Section~\ref{ModelComparison} and Fig.~\ref{FigCMD}) provides guidance 
on the probable relative covering fractions: the models suggest a thin cloud cover varying between 71\% and 75\% for SIMP0136 (i.e. dominantly darker spots in a lighter photosphere) and between 37\% to 50\% thin cloud cover for 2M2139. However, this comparison is
imperfect: our lightcurves are sensitive to changes in the surface covering fractions and insensitive to azimuthally symmetrically distributed photospheric features. Such features -- bands or evenly distributed small spots -- would not influence the light curves (i.e. distribution of the darker/lighter surface features), but would influence the relative photospheric covering fraction of the atmospheric models used.
We foresee that our dataset and similar datasets will be modeled in the near future by {\em simultaneously} fitting the mean spectra, the
spectral changes and the lightcurve shapes in a self-consistent manner, instead of the three--step procedure we followed here. 

\bibliographystyle{aa}       
\bibliography{bdrefs}   

\end{document}